\documentclass[aip,jcp,graphicx,reprint]{revtex4-1}
\usepackage{graphicx} 
\usepackage{amsmath}
\usepackage{multirow}
\usepackage{booktabs}

\begin{document}

\title{Accuracy of Vibrational Properties of Machine-Learning Force Fields for Molecular Crystal Transformations} 

\author{Natalia Goncharova}
\author{Oskar Schweighofer}
\author{Johannes Hoja}
\email[]{johannes.hoja@uni-graz.at}
\affiliation{Department of Chemistry, University of Graz, Heinrichstraße 28/IV, 8010 Graz, Austria.}

\date{9 October 2026}

\begin{abstract}
Machine-learning force fields (MLFFs) enable the calculation of transformation pathways in molecular crystals at a fraction of the computational cost of density functional theory (DFT). However, a good agreement in energies and structures does not necessarily imply an accurate description of vibrational properties, which are essential for characterizing transition states and determining free energies at finite temperatures. Herein, we assess the accuracy of system-specific SO3krates MLFFs trained on PBE+MBD reference data for describing vibrational properties of molecular crystal transformations. As examples, we consider a polymorph transition of 1,4-diiodobenzene and a single-crystal-to-single-crystal Diels-Alder reaction. The transition state structures obtained from solid-state nudged-elastic-band (NEB) calculations with the MLFFs exhibit a single imaginary $\Gamma$-point frequency associated with the respective transformation. Compared with PBE+MBD, the mean absolute errors of $\Gamma$-point vibrational frequencies and relative harmonic free energies at 300 K are about 3 cm$^{-1}$ and 1 kJ/mol, respectively. Furthermore, we compare for the Diels-Alder reaction a collective transformation pathway with a sequential one, in which individual diene-dienophile pairs react successively, and demonstrate that PBE0+MBD single-point corrections can significantly affect the predicted reaction energetics. These results demonstrate that system-specific MLFFs can reproduce not only the energies and forces but also the local curvature of the potential-energy surface with high accuracy, enabling reliable calculations of harmonic vibrational free energies, providing a basis for extending the studies of molecular crystal transformations to larger supercells.
\end{abstract}


\maketitle 

\section{Introduction}
Understanding the structures and transformations of molecular crystals is important for the pharmaceutical and electronics industries.\cite{Beran2023, Jiang2019, Wang2018, Beran2016} Recent advances in machine-learning force fields (MLFFs)\cite{Chmiela2017,Schtt2017,Chmiela2018,Schutt21a,Poltavsky2021,Unke2021,Sauceda2022,Batzner2022,Batatia2022,frank2022so3krates,frank2024euclidean,Kovcs2025} have accelerated studies of molecular crystals\cite{Zugec2024, Mladineo2024,Zhao2025,Nickerson2025} and made it feasible to investigate processes that are difficult to study with periodic first-principles calculations. One example is the study of transformation pathways and their corresponding transition states in molecular crystals. 
Nudged elastic band (NEB) calculations provide a route to these pathways in molecular and periodic systems.\cite{Henkelman2000,Gelli2026,Rohling2018} Solid-state formulations additionally allow changes in both atomic positions and lattice vectors.\cite{Sheppard2012}

Such transformations include transitions between molecular crystal polymorphs and solid-state chemical reactions. 
Controlling the crystalline form is a crucial issue, especially in pharmaceutical development, since different crystal structures of the same compound often exhibit different physical properties.\cite{Shi2022, Haleblian1969, Lee2011, Braga2022}
Crystal structure prediction methods have made substantial progress in generating and ranking possible crystal structures,\cite{Hunnisett2024,Hunnisett2024b, Zhou2025,Firaha2023} although the resulting polymorphic energy landscapes often contain more low-energy structures than are observed experimentally.\cite{Butler2023} 
Understanding their accessibility therefore also requires information about barriers connecting them. Solid-state chemical reactions can proceed as single-crystal-to-single-crystal transformations. One example is the Diels-Alder reaction considered here, in which the reactants are initially arranged in a co-crystal.\cite{Khorasani2015}  

Electronic energies alone do not account for vibrational contributions to relative stabilities and activation barriers at finite temperatures. The energy differences between molecular crystal structures can be very small, placing demanding requirements on the treatment of intermolecular interactions. Furthermore, zero-point motion and thermal vibrations can modify polymorph stability rankings,\cite{Nyman2015, Hoja2019, Weatherby2022, Kapil2022, Nyman2016} with low-frequency intermolecular modes making crucial contributions to relative vibrational free energies.  Harmonic phonon calculations provide an accessible first step for including such contributions. However, thermal expansion and other anharmonic effects can introduce further changes in stability.\cite{Raimbault2019b, Dolgonos2019,Hoja2019,Hattori2024} Making such calculations practical requires efficient force evaluations for larger systems.

Machine-learning approaches have already been successfully applied to molecular crystal energy landscapes and vibrations.\cite{Musil2018,Raimbault2019,Krynski2021,DellaPia2025,Gurlek2025} Transferable MLFFs such as SO3LR\cite{Kabylda2025} or MACE-OFF\cite{Kovcs2025} provide pretrained potentials suitable for a variety of systems. However, the accuracy of these models for a particular application depends on the underlying reference method and the configurations represented in training. Fine-tuning can improve the accuracy for selected systems by incorporating additional system-specific reference data.\cite{Kaur2025,DellaPia2025} 
 
In a previous publication,\cite{Goncharova2025} we trained system-specific MLFFs utilizing the SO3krates\cite{frank2022so3krates,frank2024euclidean} neural network architecture solely on periodic PBE+MBD reference data. This enabled a direct comparison of MLFF results with those from corresponding calculations utilizing exactly the reference method (PBE+MBD). The resulting MLFF structures and relative electronic energies along NEB pathways were in excellent agreement with PBE+MBD results, with a mean absolute error in relative lattice energies of 0.4 kJ/mol. 

Herein, we evaluate the accuracy of these MLFFs for harmonic vibrational properties, which depend on second derivatives of energies w.r.t atomic positions. Since the MLFFs were trained on energies and their first derivatives (forces), the accuracy of vibrations requires a separate assessment. Having access to MLFF vibrational frequencies enables us to check the validity of transition states obtained from NEB calculations and to calculate harmonic free energies at room temperature. Specifically, we compare MLFF harmonic vibrational frequencies and normal modes, as well as relative vibrational and Helmholtz free energies at 300 K, with PBE+MBD results for local minima and transition states. As representative examples, we utilize the polymorph transition between forms $\alpha$ and $\beta$ of 1,4-diiodobenzene and a single-crystal-to-single-crystal Diels-Alder reaction starting from a co-crystal of bis(N-allylimino)-1,4-dithiin and 9-bromoanthracene. We show that these MLFFs accurately reproduce the harmonic vibrational properties obtained with PBE+MBD and provide a basis for extending these calculations to larger periodic systems.

Finally, we investigate different pathway types for the Diels-Alder reaction. In addition to a collective pathway, in which all four diene-dienophile moieties in the unit cell react simultaneously, we also examine a sequential pathway in which the diene-dienophile moieties react successively. We also evaluate PBE0+MBD single-point corrections along these pathways to assess the sensitivity of the reaction energetics to the used density functional approximation. The sequential pathway provides a basis for subsequent studies of localized reactions in larger supercells.

\section{Computational Methods}

We investigate the polymorph transition from the $\alpha$ to the $\beta$ polymorph of 1,4-diiodobenzene\cite{Alcob1994} and a single-crystal-to-single-crystal Diels-Alder reaction of bis(N-allylimino)-1,4-dithiin with 9-bromoanthracene\cite{Khorasani2015}. As in our previous work\cite{Goncharova2025}, which focused on MLFF training and initial NEB pathways, the CSD reference codes of the experimental structures are ZZZPRO05 and ZZZPRO07 for the two polymorphs, and FUSCIH and FURROB for the co-crystal and Diels-Alder product, respectively.\cite{Khorasani2015}

Unless noted otherwise, DFT calculations were performed with  FHI-aims \cite{Blum2009,Knuth2015,Ren2012,Yu2018,Havu2009,Ihrig2015} (version 231212) utilizing PBE+MBD-NL\cite{Perdew1996,Tkatchenko2012, Ambrosetti2014, Hermann2020}  with the 2020 light species default settings for basis functions and integration grids. Additionally, selected single-point calculations were performed using PBE0+MBD-NL\cite{Adamo1999} with light species default settings. The number of k-points ($n$) in each direction was chosen to satisfy $n \times l \geq 18$ \AA{} or 30 \AA{} for the DFT and MBD k-point grids, respectively, with $l$ being the cell length in the respective direction. The convergence criteria were defined as $10^{-6}$ eV for the total energy, 10$^{-5}$ electrons per \AA{}$^{3}$ for the charge density, 10$^{-4}$ eV/\AA{} for forces, and 10$^{-3}$ eV for the sum of the eigenvalues. Geometry optimizations were converged to a maximum force component of 0.005 eV/\AA{}. The Atomic Simulation Environment (ASE)\cite{ase-paper} was utilized for all calculations. 

\begin{figure*}
\includegraphics[width=\textwidth]{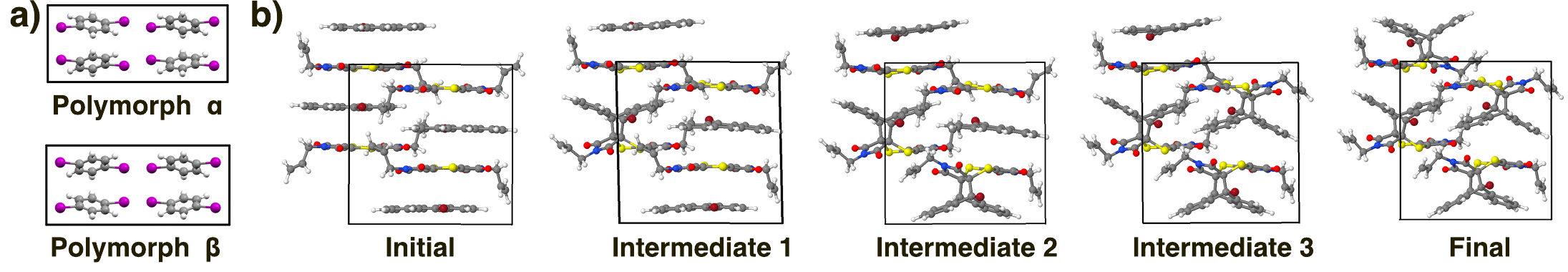}
\caption{\label{fig1} SO3krates MLFF-optimized structures of $\alpha$ and $\beta$ forms of 1,4-diiodobenzene (a) along with the initial co-crystal, final product, and the three intermediate structures along the sequential, four-step Diels-Alder reaction pathway (b).}
\end{figure*}

Furthermore, we utilize our previously trained, system-specific  SO3krates MLFFs.\cite{frank2022so3krates,frank2024euclidean} For the training PBE+MBD/light reference data obtained from several very short molecular dynamics trajectories and lattice relaxations of interpolated pathways were used (see Ref. \citenum{Goncharova2025} for details). 

To determine transition-state structures, nudged elastic band  (NEB) calculations were carried out with the solid-state NEB code (SSNEB), which is part of the transition state library for ASE (TSASE),\cite{Sheppard2012} utilizing the climbing image method.\cite{Henkelman2000} Typically, 15 images were utilized for 1,4-diiodobenzene and 30 for the Diels-Alder reaction in each calculation. The determination of the transition state of a reaction generally involves several steps: first, entire NEB pathways were optimized, as discussed in our previous publication\cite{Goncharova2025}; then, the part of the pathway surrounding the image with the highest energy was extracted and 
further refined, when necessary. During this refinement, only the climbing image was allowed to move.

For the harmonic free energy calculations we used the phonopy code\cite{Togo2015, phonopy-phono3py-JPCM} (version 3.1.0) with atomic displacements of 0.005~\AA~and a $q$-point grid satisfying the condition $q \cdot x > 50$~\AA, with $x$ being the cell length in that direction. For the finite-displacement calculations supercells with lengths of at least 12~\AA~ in each direction were used. To compare the modes obtained with PBE+MBD/light and the SO3krates MLFF, the absolute inner product of each pair of modes between structures was calculated, yielding 1 for identical modes and 0 for orthogonal modes. The optimal mode correspondence between two structures was then determined by maximizing the sum of overlaps $O$ between all modes, which essentially reduces to a two-dimensional assignment problem\cite{Crouse2016},
\begin{equation}
 O= \sum^{N}_{i}|\langle\mathbf{q}_{i},\mathbf{q}_{j(i)}\rangle|
\end{equation} 
where $N$ is the total number of modes, $\mathbf{q}_{i}$ is a normalized mode (vector) of displacement $i$ of the first structure, and $\mathbf{q}_{j(i)}$ is the normalized vector of displacement $j$ of the second structure, such that the sum is maximized.

We also compared pathways in which the four diene-dienophile moieties in the unit cell react simultaneously or one after another. 
To investigate these transitions SSNEB calculations with the system-specific SO3krates MLFF were performed. First,  intermediate structures for each partial reaction step were generated and optimized.
The interpolated pathways were generated with our previously described code, which was further improved to reduce the number of required input parameters to the minimum necessary for the interpolation and to ensure that manual fragment definition is required only in more complex cases. Unlike our previous work, where treatment of the co-crystal of 9-bromoanthracene and bis(N-allylimino)-1,4-dithiin required in total ten fragments, the current method uses only four. Specifically bis(N-allylimino)-1,4-dithiin is partitioned into two propenyl groups and the remaining central fragment. 
The current algorithm continues to use spherical linear interpolation (SLERP)\cite{Shoemake1985} for the rotation of fragments, although now as part of an as-rigid-as-possible (ARAP) algorithm.\cite{Sorkine} Our implementation draws upon the protocol described in Refs. \citenum{Nguyen2017, Nguyen2018}, adapted to incorporate periodic boundary conditions constraints. The used cell interpolation combines SLERP for cell rotation with linear scaling of the cell parameters.

\section{Results and Discussion}

\subsection{Transition-state geometries}
First, we evaluate the quality of transition-state geometries calculated with our SO3krates MLFF. For this comparison we utilize two transformations, for which we can still perform full PBE+MBD reference calculations: a) the polymorph transition of 1,4-diiodobenzene from form $\alpha$ to form $\beta$ and b) a single-crystal-to-single-crystal Diels-Alder reaction of bis(N-allylimino)-1,4-dithiin with 9-bromoanthracene (see Fig. \ref{fig1}).
For the latter we use for this benchmark a reaction pathway, where only one out of the four diene-dienophile moieties reacts (reaction from initial co-crystal structure to our intermediate 1 structure), since this could also still be calculated fully at the DFT level. At this stage our objective is to evaluate the performance of our MLFF compared to the PBE+MBD reference and we do not imply that this particular diene-dienophile moiety in a crystal is more likely to react as the first one.

For the SO3krates MLFF, our climbing image NEB calculations provided directly transition-state structures with exactly one imaginary $\Gamma$-point vibrational frequency, with the corresponding mode describing the correct transformation motion, as required for a proper transition state. The PBE+MBD NEB calculation for 1,4-diiodobenzene yields at first an approximate transition state with two imaginary frequencies. In our previous publication\cite{Goncharova2025}, we were able to remove this additional imaginary frequency via small displacements of atoms along that particular mode. Now, we optimize the DFT transition state via an additional NEB calculation that restricts movement to the climbing image as described above, yielding then a proper transition state. 
To identify the PBE+MBD transition state for the Diels-Alder reaction (to our intermediate 1 structure), we extracted a segment of the corresponding MLFF pathway, surrounding the potential transition state and optimized this pathway with PBE+MBD, allowing only the image with the highest energy to move. The final transition state structure was verified as previously with MLFF by calculating phonon frequencies at the $\Gamma$-point.

\begin{figure*}
\includegraphics[width=0.8\textwidth]{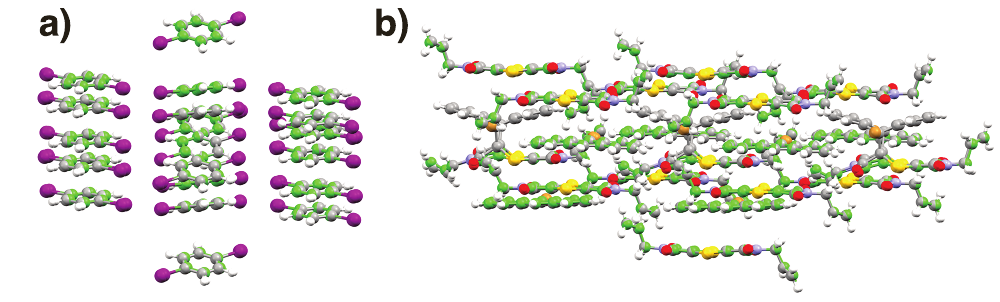}
\caption{\label{fig2} Overlay of the 1,4-diiodobenzene transition state structures  (a) and the Diels-Alder transition state structures for a reaction to our intermediate 1 structure (b) obtained with PBE+MBD and our SO3krates MLFF.}
\end{figure*}

The resulting SO3krates MLFF and PBE+MBD transition-state structures are visualized and compared in Fig. \ref{fig2} and Table \ref{tab1} shows a comparison of obtained SO3krates MLFF and PBE+MBD structures in terms of the $\mathrm{RMSD}_{20}$, which is the root-mean-squared deviation for a cluster containing 20 molecules. Visually the obtained transition state structures are virtually identical and both corresponding $\mathrm{RMSD}_{20}$ values are less than 0.14 \AA, highlighting an excellent agreement between the PBE+MBD and SO3krates MLFF transition-state structures. When comparing structures for the different obtained local minima, all $\mathrm{RMSD}_{20}$ values are smaller than 0.24 \AA{}, with the worst agreement being the intermediate 1 structure for the Diels-Alder reaction.

\begin{table}[h]
\setlength{\tabcolsep}{6.5pt}
\centering
\caption{\label{tab1}RMSD$_{20}$ values between PBE+MBD and  SO3krates MLFF structures in \AA{}.}
\label{tab1}
\begin{tabular}{@{}lrrrr@{}}
\hline\hline
 Structure	&	Initial & TS & Intermediate 1 & Final\\
\hline
p-Diiodobenzene & 0.134 & 0.137 & n/a & 0.186 \\
Diels-Alder & 0.117 & 0.130 & 0.238 & 0.097 \\

\hline\hline
\end{tabular}\\
\end{table}

\subsection{Vibrational modes and frequencies}

Next, we evaluate the quality of the MLFF vibrational modes at the $\Gamma$-point and their corresponding vibrational frequencies by comparing the SO3krates MLFF results with PBE+MBD data for local minima and obtained transition states. Since the ordering of vibrational modes is not necessarily the same in different methods, we sorted the modes based on their maximal overlap as described in the Computational Methods section. Table \ref{tab2} shows the mean error (ME), the mean absolute error (MAE), and the maximal absolute error (Max. error) of the SO3krates MLFF compared to PBE+MBD for the 1,4-diiodobenzene structures. The ME is across the structures very consistent and varies between -0.6 and 0.5 cm$^{-1}$, yielding an overall value of only 0.1 cm$^{-1}$. The MAE varies across the three structures from 2.9 to 3.6 cm$^{-1}$ with an overall value of 3.2 cm$^{-1}$. If we restrict this evaluation only to low-frequency modes below 300 cm$^{-1}$ the overall MAE slightly increases to about 5~cm$^{-1}$. This shows that high-frequency modes are with the MLFF better described than low-frequency modes, which is not surprising given the quite collective and mostly intermolecular motion of the latter.  Overall the MLFF yields very accurate vibrational frequencies. However, some individual frequencies can differ significantly more with the maximal absolute error among the local minima being 20 cm$^{-1}$. For the transition state one mode even differs by 32 cm$^{-1}$. 

\begin{figure}
\includegraphics[width=\columnwidth]{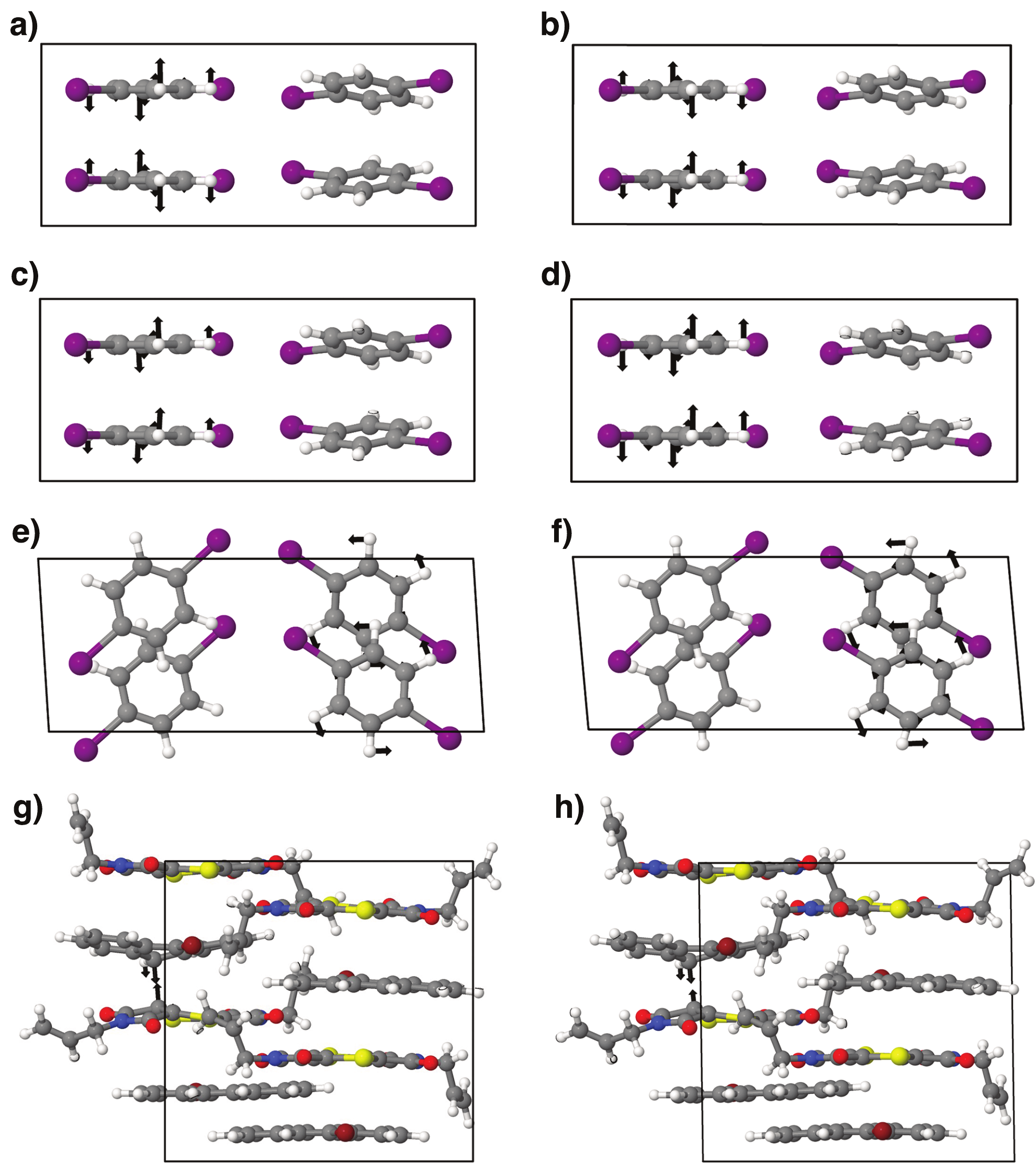}
\caption{\label{fig3} Comparison of $\Gamma$-point vibrational modes: imaginary mode of the transition state during the 1,4-diiodobenzene transition from the $\alpha$ to the $\beta$ polymorph calculated with PBE+MBD (a) and SO3krates MLFF (b), mode with the maximal deviation in frequency for 1,4-diiodobenzene (31.7 cm$^{-1}$) with PBE+MBD (c) and SO3krates MLFF (d), minimal overlap mode (0.72) for 1,4-diiodobenzene with PBE+MBD (e) and SO3krates MLFF (f), imaginary mode of the Diels-Alder intermediate 1 transition state with PBE+MBD (g) and SO3krates MLFF (h). }
\end{figure}

Now we want to put these frequency differences in context of the actual motion described by the corresponding mode, which is illustrated for several examples in Fig. \ref{fig3}. The imaginary mode of the transition state (a, b) describes with both methods the rotation of the two left molecules leading to the desired polymorph transition. Only the signs of the displacements are inverted, which amounts still to the same description since the signs are arbitrary in the harmonic approximation. For the transition state the difference in the imaginary frequency between PBE+MBD and the MLFF amounts to 13.4 cm$^{-1}$. In (c, d) one can see the same comparison for the mode with the largest frequency deviation for 1,4-diiodobenzene, which amounts to 31.7 cm$^{-1}$, but both methods describe the same dominant motion. Additionally, we show in (e, f) a comparison of the mode with the smallest overlap value for 1,4-diiodobenzene, which also describes in both cases the same motion with similar frequencies.

For the local-minimum structures and the transition state for the Diels-Alder reaction we observe a fairly similar picture (see Table \ref{tab3}). Overall the ME amounts to -1.2 cm$^{-1}$ and the overall MAE to 2.8 cm$^{-1}$, which is very consistent with the results for the previously discussed polymorph transition. Interestingly, we observe here no real difference between high- and low-frequency modes, since restricting the analyses to modes with less than 300 cm$^{-1}$ yields a MAE of 2.6 cm$^{-1}$. However, the maximal absolute error is with 86.3 cm$^{-1}$ significantly larger than for 1,4-diiodobenzene. It corresponds to the imaginary mode of the transition state (see Fig. \ref{fig3}(g, h)), but also in this case the qualitative motion is identical. For local-minimum structures the maximal error is located in a similar range as for 1,4-diiodobenzene.

\begin{table}[h]
\setlength{\tabcolsep}{8pt}
\centering
\caption{\label{tab2}Frequency comparison in cm$^{-1}$ of the modes obtained at the $\Gamma$-point with PBE+MBD and the SO3krates MLFF for 1,4-diiodobenzene.}
\label{tab2}
\begin{tabular}{@{}lrrrr@{}}
\hline\hline
 Error Statistic&	$\alpha$	&	$\beta$	&	TS &  	$\alpha+\beta+$TS\\
\hline
ME	&	0.2	&	0.5 &	-0.6   &  0.1\\
MAE	&	3.0    &	2.9	&	3.6  &    3.2  \\
Max. error	&	17.7    &	19.5	&	31.7  & 31.7\\

\hline\hline
\end{tabular}\\
\end{table}

\begin{table}[h]
\setlength{\tabcolsep}{7pt}
\centering
\caption{Frequency comparison in cm$^{-1}$ of the modes obtained at the $\Gamma$-point with PBE+MBD and the SO3krates MLFF for co-crystal of bis(N-allylimino)-1,4-dithiin and 9-
bromoanthracene (Initial), the respective Diels-Alder product (Final), intermediate 1 after the reaction of the first diene-dienophile moiety (Interm. 1), a corresponding transition state (TS), and overall statistics (All).}
\label{tab3}
\begin{tabular}{@{}lrrrrr@{}}
\hline\hline
 Error Statistic&	Initial & TS & 	Interm. 1 & Final& All\\
\hline
ME	&  -0.6 &	-1.8 &  -0.6 & -1.6 & -1.2\\
MAE	& 2.5 &	3.1 &	 2.8& 2.9   &  2.8\\
Max. error	&13.5 &	86.3 &  16.6 &	12.0  & 86.3\\

\hline\hline
\end{tabular}\\
\end{table}

\subsection{Free energies}

After establishing the close agreement between the vibrational modes and frequencies obtained with our SO3krates MLFFs and PBE+MBD, we next assess how this accuracy translates to free energies at finite temperatures. Our previous study\cite{Goncharova2025} reported electronic energy profiles for the two herein considered transformations. Here, we now obtain and compare Helmholtz free energies by adding vibrational free energies, calculated with proper reciprocal space sampling, to the electronic energies.

\begin{figure}
\includegraphics[width=\columnwidth]{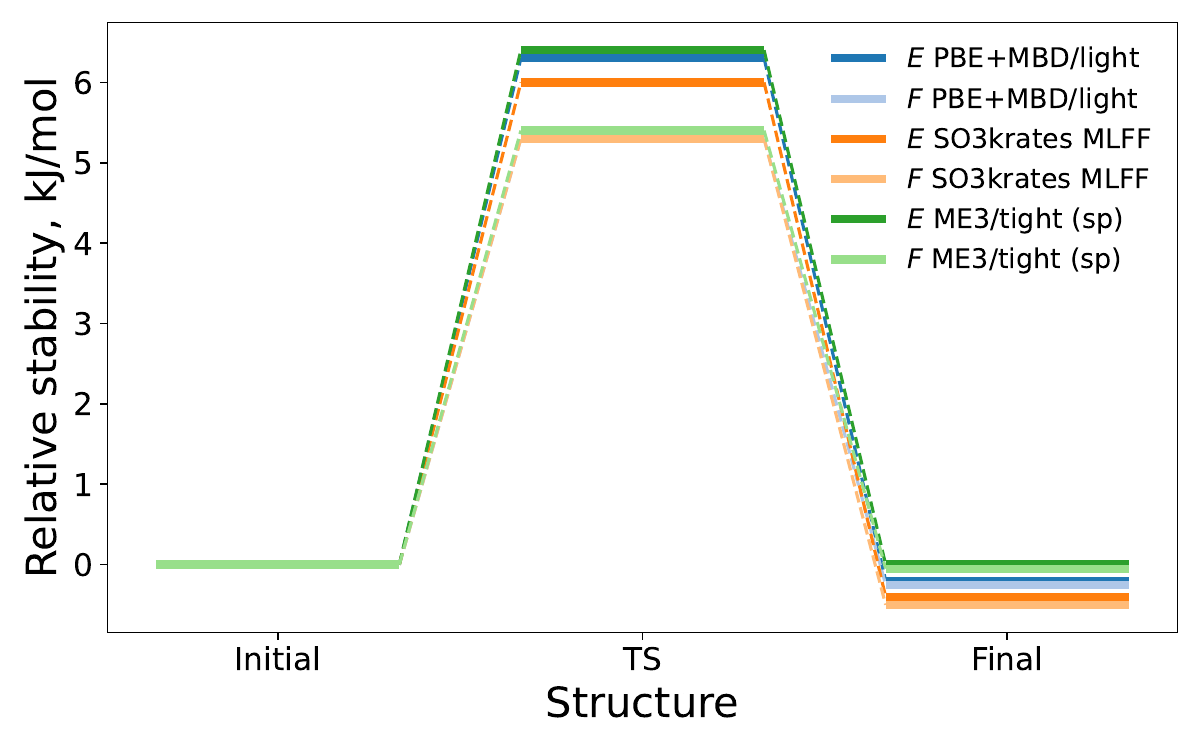}

\caption{\label{fig4} Relative electronic energies ($E$) and free energies ($F$) at 300 K in kJ/mol (normalized per molecule) for the considered 1,4-diiodobenzene polymorph transition: $\alpha$ polymorph (Initial),  transition state (TS) and $\beta$ polymorph (Final). For $F$ ME3/tight (sp) the electronic energy was calculated on this level and the vibrational free energy at PBE+MBD/light. }
\end{figure}

Table \ref{tab4} shows the resulting relative Helmholtz free energies and vibrational free energies at 300 K as well as at 0 K for 1,4-diiodobenzene. At 0 K, the vibrational contribution amounts to zero-point energy effects. For both quantities and temperatures our SO3krates MLFF agrees with the PBE+MBD reference within 0.3 kJ/mol. However, in this case vibrational free energies contribute only up to 1 kJ/mol to the shown energy differences, with the vast majority stemming from zero-point-energy effects. Electronic energies and free energies are also visualized in Fig. \ref{fig4}. It can be seen that the vibrational free energy at room temperature lowers the energy barrier by about 1 kJ/mol and the $\beta$ polymorph remains very slightly more stable than the $\alpha$ polymorph. In addition to the herein used methods, we also show energy differences with ME3/tight, which is a multimer embedding method that approximates converged periodic PBE0+MBD results.\cite{Hoja2023,List2026,Goncharova2025} In this case our PBE+MBD and MLFF energies agree very well with the ME3/tight benchmark result. 

\begin{figure}
\includegraphics[width=\columnwidth]{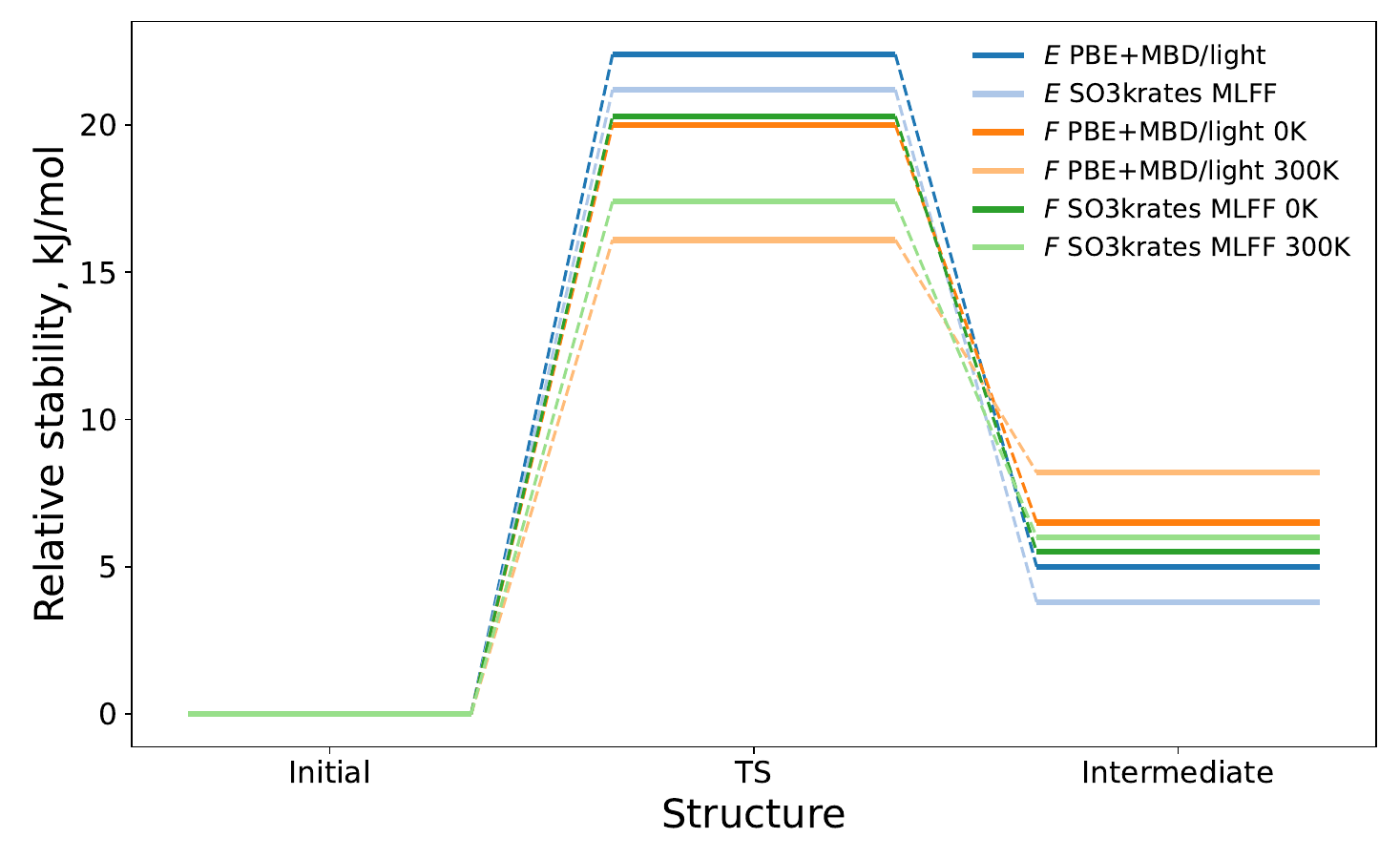}
\caption{\label{fig5} Relative electronic energies ($E$) and free energies ($F$) at 0 K and 300 K in kJ/mol (normalized per diene-dienophile moiety) for the considered Diels-Alder reaction: initial bis(N-allylimino)-1,4-dithiin and 9-bromoanthracene co-crystal (Initial), the transition state (TS) and the intermediate 1 structure (Intermediate). }
\end{figure}

The corresponding results for the Diels-Alder reaction (reaction of one diene-dienophile moiety to structure intermediate 1) are shown in Table \ref{tab5} and Fig. \ref{fig5}. Note that we normalize these energies per diene-dienophile moiety in the initial co-crystal since we utilized this normalization in our previous publication\cite{Goncharova2025} and discuss later also the fully collective reaction pathway. Therefore, the energy change of the full unit cell is a factor of four higher. The free energies of the SO3krates MLFF agree within 2.2 kJ/mol with the PBE+MBD reference, while the vibrational free energies agree within 2.5 kJ/mol. In this case, vibrational free energy effects play a much larger role and decrease the energy barrier by  6 kJ/mol at 300~K in the case of PBE+MBD. Note that our intermediate 1 structure is described as less stable than the initial co-crystal structure. This may reflect limitations of semi-local functionals like PBE in describing Diels-Alder reaction energies and will be addressed below.

\begin{table}[h]
\setlength{\tabcolsep}{6pt}
\centering
\caption{\label{tab4}Relative Helmholtz ($\Delta F$) and vibrational free energies ($\Delta F^{\mathrm{vib}}$) in kJ/mol (normalized per molecule) for 1,4-diiodobenzene at 0 K and 300 K. The energies are given relative to the $\alpha$ polymorph, where '$\beta$' stands for the energy difference between two polymorphs and 'ts' for the energy difference to the transition state.}
\label{tab4}
\begin{tabular}{@{}lrrrrr@{}}
\hline\hline
 \multirow{2}{*}{Energy} & \multicolumn{2}{c}{0 K} & & \multicolumn{2}{c}{300 K} \\
\cline{2-3} \cline{5-6}
& PBE+MBD & MLFF && PBE+MBD & MLFF \\
\hline
 $\Delta F_{\beta}$ &  -0.2   & -0.4 && -0.2 & -0.5\\
 $\Delta F^{\mathrm{vib}}_{\beta}$  &	0.0    &	-0.1 && 0.0 & -0.2\\
$\Delta F_{\mathrm{ts}}$ & 5.5  & 5.4    &&	5.3 & 5.3  \\
$\Delta F^{\mathrm{vib}}_{\mathrm{ts}}$ & -0.9 &  -0.7    &&	-1.0	 &	-0.8\\
\hline\hline
\end{tabular}\\
\end{table}

\begin{table}[h]
\setlength{\tabcolsep}{6pt}
\centering
\caption{\label{tab5}Relative Helmholtz ($\Delta F$) and vibrational free energies ($\Delta F^{\mathrm{vib}}$) in kJ/mol for the Diels-Alder reaction (normalized per diene-dienophile moiety) at 0 K and 300 K. The energies are given relative to the initial bis(N-allylimino)-1,4-dithiin and 9-bromoanthracene co-crystal, and the subscript denotes if the energy difference is between the transition state (ts) or the intermediate 1 structure (int) and the initial co-crystal.}
\label{tab5}
\begin{tabular}{@{}lrrrrr@{}}
\hline\hline
 \multirow{2}{*}{Energy} & \multicolumn{2}{c}{0 K} & & \multicolumn{2}{c}{300 K} \\
\cline{2-3} \cline{5-6}
& PBE+MBD & MLFF && PBE+MBD & MLFF \\
\hline
$\Delta F_{\mathrm{ts}}$  &20.0 & 20.3  && 	16.1 &	 17.4 \\
$\Delta F^{\mathrm{vib}}_{\mathrm{ts}}$ &-2.4   &  -1.0   && -6.3  &	-3.8 \\
$\Delta F_{\mathrm{int}}$ & 6.5 & 5.5 && 8.2 & 6.0 \\
$\Delta F^{\mathrm{vib}}_{\mathrm{int}}$ & 1.6 & 1.7 && 3.3 & 2.2 \\ 
\hline\hline
\end{tabular}\\
\end{table}

\subsection{Collective vs. sequential reaction pathways}
\begin{figure*}
\includegraphics[width=\textwidth]{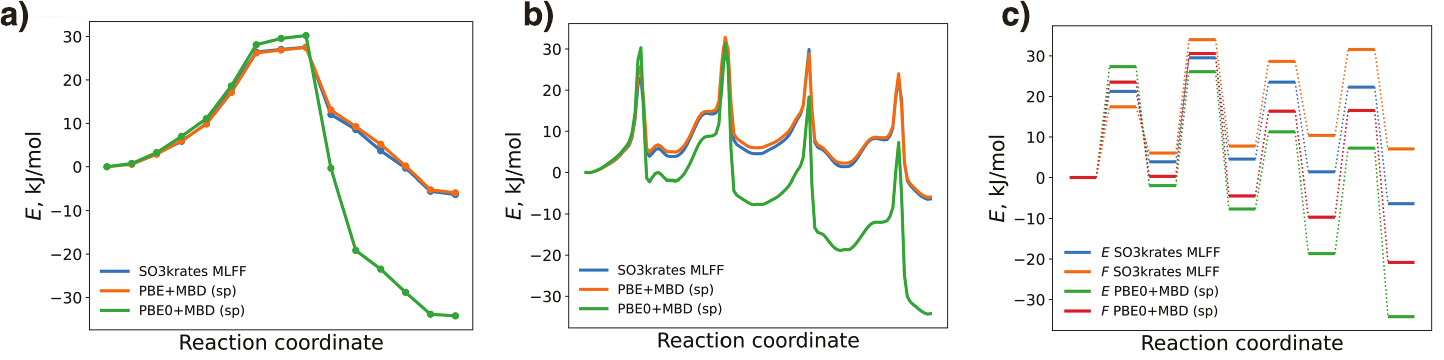}
\caption{\label{fig6} Energy profiles of potential pathways for the Diels-Alder reaction during transformations from the initial co-crystal of
bis(N-allylimino)-1,4-dithiin and 9-bromoanthracene to the final product optimized with the SO3krates MLFF and including PBE+MBD/light and PBE0+MBD/light single-point energies calculated on the MLFF structures. The total energy profile of (a) is minimized directly in one step, while the profile (b) is constructed by combining the NEB optimized energy profiles of each reaction step (sequential pathway), rather than through a global optimization of the entire reaction. Both (a) and (b) only show electronic energies without any vibrational contribution. MLFF and PBE+MBD energies in (a) are from our previous study\cite{Goncharova2025}. Part (c) shows electronic energies as well as Helmholtz free energies at 300 K for intermediates and transition states of the sequential reaction pathway from (b). The shown transition state energies in (c) correspond to structures closest to a local maximum in (b), which have exactly one imaginary $\Gamma$-point frequency. The PBE0+MBD/light free energy in (c) consists of PBE0+MBD/light electronic energies plus the MLFF vibrational free energies.}
\end{figure*}

Previously\cite{Goncharova2025}, we studied the discussed Diels-Alder reaction with NEB via a collective pathway, in which all four diene-dienophile moieties in the unit cell react at the same time (see Fig. \ref{fig6} a). This pathway has a PBE+MBD energy barrier of 27.5 kJ/mol and involves rotations of the propenyl groups. Such a collective transformation imposes synchronous motion under periodic boundary conditions, whereas the reaction may instead initiate locally and propagate through the crystal.  Therefore, we have also investigated sequential reaction pathways, in which the individual diene-dienophile moieties react successively. Fig. \ref{fig6} b) shows one representative sequential pathway connecting the initial co-crystal and the final product via three partially reacted intermediate structures (see Fig. 1). For this pathway PBE+MBD and our SO3krates MLFF give again very similar energy profiles with four distinct maxima. For all four reaction steps we obtained with the SO3krates MLFF proper transition-state structures. Note that we again normalize all energies per diene-dienophile moiety in the initial co-crystal structure, corresponding to the unit-cell energy difference divided by four. 
The highest relative electronic energy along this sequential pathway is similar to that of the collective pathway, making sequential conversion a plausible alternative within the present unit-cell model. The individual energy barriers differ along the sequence, although establishing a preferred reaction order would require a comparison of all alternative pathways.

The good correlation of PBE+MBD with our MLFFs for studying vibrational free energies and transition state structures enables us now to efficiently study this pathway of four sequential Diels-Alder reactions including vibrational contributions. Fig. \ref{fig6} c)  illustrates the corresponding reaction pathway, where energy maxima denote the transition states of each partial reaction, and minima represent initial, intermediate, and final structures. Incorporating vibrational free energy increases the relative energy for most structures by 2 to 13 kJ/mol, with the exception of the first transition state structure. Although this vibrational contribution is largest for the final product, the final product remains lower in free energy than the two previous intermediates. 

However, the final product has a positive relative Helmholtz free energy at 300 K and is therefore described as less stable than the initial co-crystal structure. This may reflect limitations of semi-local density functionals like PBE for describing Diels-Alder reaction energies\cite{Mezei2015}, motivating a comparison with the hybrid functional PBE0.  While replacing PBE with PBE0 has little effect on the relative energies of the 1,4-diiodobenzene polymorph transition, the energetics of the considered Diels-Alder reaction depend substantially on this choice. 

Therefore, we have calculated single-point energies with PBE0+MBD/light on top of the obtained MLFF structures along the reaction pathways (see green lines in Fig. \ref{fig6}a,b). 
The resulting energy profiles retain similar locations of minima and maxima along the sampled pathway, while PBE0+MBD substantially stabilizes the reacted structures relative to the initial co-crystal.

The final Diels-Alder product is now by 34.8~kJ/mol more stable than the initial co-crystal structure in terms of the electronic energy. Combining these electronic energies with the MLFF harmonic vibrational free energies at 300 K yields an approximate Helmholtz free-energy profile (red lines in Fig. \ref{fig6}c), in which the final Diels-Alder product remains more stable than the initial co-crystal by 20.8~kJ/mol. The first intermediate is close to the initial co-crystal in free energy, whereas subsequent intermediates and the final product have increasingly negative relative free energies.

In order to assess the structural changes associated with replacing PBE by PBE0, we optimized the initial and final structures with PBE0+MBD/light and compared them with their PBE+MBD counterparts. In both cases the RMSD$_{20}$ amounts to less than 0.2 \AA{}. These small structural differences, together with the similar shapes of the energy profiles, support the use of PBE0+MBD single-point corrections along the MLFF pathways for the present system.

\section{Conclusion}
We have assessed the accuracy of system-specific SO3krates MLFFs for harmonic vibrational properties of molecular crystal transformations, extending our previous benchmarks of structures and electronic energies.\cite{Goncharova2025}  For the studied 1,4-diiodobenzene polymorph transition and the single-crystal Diels-Alder reaction, the $\Gamma$-point vibrational frequencies agree with PBE+MBD reference calculations with a mean absolute error (MAE) of approximately 3 cm$^{-1}$. The relative Helmholtz free energies at 300 K show a MAE of about 1 kJ/mol w.r.t. PBE+MBD.
Beyond their role in calculating free energies, harmonic vibrations provide an essential check of the transition-state structures obtained from NEB pathways. The presented MLFF transition states each exhibit a single imaginary $\Gamma$-point frequency associated with the respective transformation.

The Diels-Alder reaction further illustrates the importance of the choice of electronic-structure reference method. Although the MLFF closely reproduces PBE+MBD results, replacing PBE with PBE0 significantly changes the reaction energetics. Combining PBE0+MBD single-point energies with MLFF vibrational contributions places the final product 21 kJ/mol below the initial co-crystal in terms of the Helmholtz free energy at 300 K. 

In addition to the collective Diels-Alder pathway, we identified a sequential pathway in which the four diene-dienophile moieties react successively via three partially reacted intermediates. 
Its highest relative electronic energy is similar to that of the collective pathway, providing an alternative description of the transformation within the unit-cell model. The demonstrated accuracy for vibrational properties and free energies provides a basis for extending these calculations to larger supercells. Such calculations would allow us to study isolated reaction events and how neighboring reacted moieties affect individual energy barriers. This could clarify whether an initial reaction event facilitates further conversion, eventually addressing nucleation and growth for molecular crystal transformations.

\begin{acknowledgments}
The computational results have been achieved in part
using the Austrian Scientific Computing (ASC) infrastructure as well as computational resources provided by the University of Graz.\\
\end{acknowledgments}


\begin{thebibliography}{73}%
\makeatletter
\providecommand \@ifxundefined [1]{%
 \@ifx{#1\undefined}
}%
\providecommand \@ifnum [1]{%
 \ifnum #1\expandafter \@firstoftwo
 \else \expandafter \@secondoftwo
 \fi
}%
\providecommand \@ifx [1]{%
 \ifx #1\expandafter \@firstoftwo
 \else \expandafter \@secondoftwo
 \fi
}%
\providecommand \natexlab [1]{#1}%
\providecommand \enquote  [1]{``#1''}%
\providecommand \bibnamefont  [1]{#1}%
\providecommand \bibfnamefont [1]{#1}%
\providecommand \citenamefont [1]{#1}%
\providecommand \href@noop [0]{\@secondoftwo}%
\providecommand \href [0]{\begingroup \@sanitize@url \@href}%
\providecommand \@href[1]{\@@startlink{#1}\@@href}%
\providecommand \@@href[1]{\endgroup#1\@@endlink}%
\providecommand \@sanitize@url [0]{\catcode `\\12\catcode `\$12\catcode `\&12\catcode `\#12\catcode `\^12\catcode `\_12\catcode `\%12\relax}%
\providecommand \@@startlink[1]{}%
\providecommand \@@endlink[0]{}%
\providecommand \url  [0]{\begingroup\@sanitize@url \@url }%
\providecommand \@url [1]{\endgroup\@href {#1}{\urlprefix }}%
\providecommand \urlprefix  [0]{URL }%
\providecommand \Eprint [0]{\href }%
\providecommand \doibase [0]{http://dx.doi.org/}%
\providecommand \selectlanguage [0]{\@gobble}%
\providecommand \bibinfo  [0]{\@secondoftwo}%
\providecommand \bibfield  [0]{\@secondoftwo}%
\providecommand \translation [1]{[#1]}%
\providecommand \BibitemOpen [0]{}%
\providecommand \bibitemStop [0]{}%
\providecommand \bibitemNoStop [0]{.\EOS\space}%
\providecommand \EOS [0]{\spacefactor3000\relax}%
\providecommand \BibitemShut  [1]{\csname bibitem#1\endcsname}%
\let\auto@bib@innerbib\@empty
\bibitem [{\citenamefont {Beran}(2023)}]{Beran2023}%
  \BibitemOpen
  \bibfield  {author} {\bibinfo {author} {\bibfnamefont {G.~J.~O.}\ \bibnamefont {Beran}},\ }\href {\doibase 10.1039/d3sc03903j} {\bibfield  {journal} {\bibinfo  {journal} {Chem. Sci.}\ }\textbf {\bibinfo {volume} {14}},\ \bibinfo {pages} {13290–13312} (\bibinfo {year} {2023})}\BibitemShut {NoStop}%
\bibitem [{\citenamefont {Jiang}\ and\ \citenamefont {Hu}(2020)}]{Jiang2019}%
  \BibitemOpen
  \bibfield  {author} {\bibinfo {author} {\bibfnamefont {H.}~\bibnamefont {Jiang}}\ and\ \bibinfo {author} {\bibfnamefont {W.}~\bibnamefont {Hu}},\ }\href {\doibase 10.1002/anie.201814439} {\bibfield  {journal} {\bibinfo  {journal} {Angew. Chem. Int. Ed.}\ }\textbf {\bibinfo {volume} {59}},\ \bibinfo {pages} {1408–1428} (\bibinfo {year} {2020})}\BibitemShut {NoStop}%
\bibitem [{\citenamefont {Wang}\ \emph {et~al.}(2018)\citenamefont {Wang}, \citenamefont {Dong}, \citenamefont {Jiang},\ and\ \citenamefont {Hu}}]{Wang2018}%
  \BibitemOpen
  \bibfield  {author} {\bibinfo {author} {\bibfnamefont {C.}~\bibnamefont {Wang}}, \bibinfo {author} {\bibfnamefont {H.}~\bibnamefont {Dong}}, \bibinfo {author} {\bibfnamefont {L.}~\bibnamefont {Jiang}}, \ and\ \bibinfo {author} {\bibfnamefont {W.}~\bibnamefont {Hu}},\ }\href {\doibase 10.1039/c7cs00490g} {\bibfield  {journal} {\bibinfo  {journal} {Chem. Soc. Rev.}\ }\textbf {\bibinfo {volume} {47}},\ \bibinfo {pages} {422–500} (\bibinfo {year} {2018})}\BibitemShut {NoStop}%
\bibitem [{\citenamefont {Beran}(2016)}]{Beran2016}%
  \BibitemOpen
  \bibfield  {author} {\bibinfo {author} {\bibfnamefont {G.~J.~O.}\ \bibnamefont {Beran}},\ }\href {\doibase 10.1021/acs.chemrev.5b00648} {\bibfield  {journal} {\bibinfo  {journal} {Chem. Rev.}\ }\textbf {\bibinfo {volume} {116}},\ \bibinfo {pages} {5567–5613} (\bibinfo {year} {2016})}\BibitemShut {NoStop}%
\bibitem [{\citenamefont {Chmiela}\ \emph {et~al.}(2017)\citenamefont {Chmiela}, \citenamefont {Tkatchenko}, \citenamefont {Sauceda}, \citenamefont {Poltavsky}, \citenamefont {Sch\"{u}tt},\ and\ \citenamefont {M\"{u}ller}}]{Chmiela2017}%
  \BibitemOpen
  \bibfield  {author} {\bibinfo {author} {\bibfnamefont {S.}~\bibnamefont {Chmiela}}, \bibinfo {author} {\bibfnamefont {A.}~\bibnamefont {Tkatchenko}}, \bibinfo {author} {\bibfnamefont {H.~E.}\ \bibnamefont {Sauceda}}, \bibinfo {author} {\bibfnamefont {I.}~\bibnamefont {Poltavsky}}, \bibinfo {author} {\bibfnamefont {K.~T.}\ \bibnamefont {Sch\"{u}tt}}, \ and\ \bibinfo {author} {\bibfnamefont {K.-R.}\ \bibnamefont {M\"{u}ller}},\ }\href {\doibase 10.1126/sciadv.1603015} {\bibfield  {journal} {\bibinfo  {journal} {Sci. Adv.}\ }\textbf {\bibinfo {volume} {3}},\ \bibinfo {pages} {e1603015} (\bibinfo {year} {2017})}\BibitemShut {NoStop}%
\bibitem [{\citenamefont {Sch\"{u}tt}\ \emph {et~al.}(2017)\citenamefont {Sch\"{u}tt}, \citenamefont {Arbabzadah}, \citenamefont {Chmiela}, \citenamefont {M\"{u}ller},\ and\ \citenamefont {Tkatchenko}}]{Schtt2017}%
  \BibitemOpen
  \bibfield  {author} {\bibinfo {author} {\bibfnamefont {K.~T.}\ \bibnamefont {Sch\"{u}tt}}, \bibinfo {author} {\bibfnamefont {F.}~\bibnamefont {Arbabzadah}}, \bibinfo {author} {\bibfnamefont {S.}~\bibnamefont {Chmiela}}, \bibinfo {author} {\bibfnamefont {K.~R.}\ \bibnamefont {M\"{u}ller}}, \ and\ \bibinfo {author} {\bibfnamefont {A.}~\bibnamefont {Tkatchenko}},\ }\href {\doibase 10.1038/ncomms13890} {\bibfield  {journal} {\bibinfo  {journal} {Nat. Commun.}\ }\textbf {\bibinfo {volume} {8}},\ \bibinfo {pages} {13890} (\bibinfo {year} {2017})}\BibitemShut {NoStop}%
\bibitem [{\citenamefont {Chmiela}\ \emph {et~al.}(2018)\citenamefont {Chmiela}, \citenamefont {Sauceda}, \citenamefont {M\"{u}ller},\ and\ \citenamefont {Tkatchenko}}]{Chmiela2018}%
  \BibitemOpen
  \bibfield  {author} {\bibinfo {author} {\bibfnamefont {S.}~\bibnamefont {Chmiela}}, \bibinfo {author} {\bibfnamefont {H.~E.}\ \bibnamefont {Sauceda}}, \bibinfo {author} {\bibfnamefont {K.-R.}\ \bibnamefont {M\"{u}ller}}, \ and\ \bibinfo {author} {\bibfnamefont {A.}~\bibnamefont {Tkatchenko}},\ }\href {\doibase 10.1038/s41467-018-06169-2} {\bibfield  {journal} {\bibinfo  {journal} {Nat. Commun.}\ }\textbf {\bibinfo {volume} {9}},\ \bibinfo {pages} {3887} (\bibinfo {year} {2018})}\BibitemShut {NoStop}%
\bibitem [{\citenamefont {Sch{\"u}tt}, \citenamefont {Unke},\ and\ \citenamefont {Gastegger}(2021)}]{Schutt21a}%
  \BibitemOpen
  \bibfield  {author} {\bibinfo {author} {\bibfnamefont {K.}~\bibnamefont {Sch{\"u}tt}}, \bibinfo {author} {\bibfnamefont {O.}~\bibnamefont {Unke}}, \ and\ \bibinfo {author} {\bibfnamefont {M.}~\bibnamefont {Gastegger}},\ }in\ \href {https://proceedings.mlr.press/v139/schutt21a.html} {\emph {\bibinfo {booktitle} {Proceedings of the 38th International Conference on Machine Learning}}},\ \bibinfo {series} {Proceedings of Machine Learning Research}, Vol.\ \bibinfo {volume} {139},\ \bibinfo {editor} {edited by\ \bibinfo {editor} {\bibfnamefont {M.}~\bibnamefont {Meila}}\ and\ \bibinfo {editor} {\bibfnamefont {T.}~\bibnamefont {Zhang}}}\ (\bibinfo  {publisher} {PMLR},\ \bibinfo {year} {2021})\ pp.\ \bibinfo {pages} {9377--9388}\BibitemShut {NoStop}%
\bibitem [{\citenamefont {Poltavsky}\ and\ \citenamefont {Tkatchenko}(2021)}]{Poltavsky2021}%
  \BibitemOpen
  \bibfield  {author} {\bibinfo {author} {\bibfnamefont {I.}~\bibnamefont {Poltavsky}}\ and\ \bibinfo {author} {\bibfnamefont {A.}~\bibnamefont {Tkatchenko}},\ }\href {\doibase 10.1021/acs.jpclett.1c01204} {\bibfield  {journal} {\bibinfo  {journal} {J. Phys. Chem. Lett.}\ }\textbf {\bibinfo {volume} {12}},\ \bibinfo {pages} {6551–6564} (\bibinfo {year} {2021})}\BibitemShut {NoStop}%
\bibitem [{\citenamefont {Unke}\ \emph {et~al.}(2021)\citenamefont {Unke}, \citenamefont {Chmiela}, \citenamefont {Gastegger}, \citenamefont {Sch\"{u}tt}, \citenamefont {Sauceda},\ and\ \citenamefont {M\"{u}ller}}]{Unke2021}%
  \BibitemOpen
  \bibfield  {author} {\bibinfo {author} {\bibfnamefont {O.~T.}\ \bibnamefont {Unke}}, \bibinfo {author} {\bibfnamefont {S.}~\bibnamefont {Chmiela}}, \bibinfo {author} {\bibfnamefont {M.}~\bibnamefont {Gastegger}}, \bibinfo {author} {\bibfnamefont {K.~T.}\ \bibnamefont {Sch\"{u}tt}}, \bibinfo {author} {\bibfnamefont {H.~E.}\ \bibnamefont {Sauceda}}, \ and\ \bibinfo {author} {\bibfnamefont {K.-R.}\ \bibnamefont {M\"{u}ller}},\ }\href {\doibase 10.1038/s41467-021-27504-0} {\bibfield  {journal} {\bibinfo  {journal} {Nat. Commun.}\ }\textbf {\bibinfo {volume} {12}},\ \bibinfo {pages} {7273} (\bibinfo {year} {2021})}\BibitemShut {NoStop}%
\bibitem [{\citenamefont {Sauceda}\ \emph {et~al.}(2022)\citenamefont {Sauceda}, \citenamefont {Gálvez-González}, \citenamefont {Chmiela}, \citenamefont {Paz-Borbón}, \citenamefont {M\"{u}ller},\ and\ \citenamefont {Tkatchenko}}]{Sauceda2022}%
  \BibitemOpen
  \bibfield  {author} {\bibinfo {author} {\bibfnamefont {H.~E.}\ \bibnamefont {Sauceda}}, \bibinfo {author} {\bibfnamefont {L.~E.}\ \bibnamefont {Gálvez-González}}, \bibinfo {author} {\bibfnamefont {S.}~\bibnamefont {Chmiela}}, \bibinfo {author} {\bibfnamefont {L.~O.}\ \bibnamefont {Paz-Borbón}}, \bibinfo {author} {\bibfnamefont {K.-R.}\ \bibnamefont {M\"{u}ller}}, \ and\ \bibinfo {author} {\bibfnamefont {A.}~\bibnamefont {Tkatchenko}},\ }\href {\doibase 10.1038/s41467-022-31093-x} {\bibfield  {journal} {\bibinfo  {journal} {Nat. Commun.}\ }\textbf {\bibinfo {volume} {13}},\ \bibinfo {pages} {3733} (\bibinfo {year} {2022})}\BibitemShut {NoStop}%
\bibitem [{\citenamefont {Batzner}\ \emph {et~al.}(2022)\citenamefont {Batzner}, \citenamefont {Musaelian}, \citenamefont {Sun}, \citenamefont {Geiger}, \citenamefont {Mailoa}, \citenamefont {Kornbluth}, \citenamefont {Molinari}, \citenamefont {Smidt},\ and\ \citenamefont {Kozinsky}}]{Batzner2022}%
  \BibitemOpen
  \bibfield  {author} {\bibinfo {author} {\bibfnamefont {S.}~\bibnamefont {Batzner}}, \bibinfo {author} {\bibfnamefont {A.}~\bibnamefont {Musaelian}}, \bibinfo {author} {\bibfnamefont {L.}~\bibnamefont {Sun}}, \bibinfo {author} {\bibfnamefont {M.}~\bibnamefont {Geiger}}, \bibinfo {author} {\bibfnamefont {J.~P.}\ \bibnamefont {Mailoa}}, \bibinfo {author} {\bibfnamefont {M.}~\bibnamefont {Kornbluth}}, \bibinfo {author} {\bibfnamefont {N.}~\bibnamefont {Molinari}}, \bibinfo {author} {\bibfnamefont {T.~E.}\ \bibnamefont {Smidt}}, \ and\ \bibinfo {author} {\bibfnamefont {B.}~\bibnamefont {Kozinsky}},\ }\href {\doibase 10.1038/s41467-022-29939-5} {\bibfield  {journal} {\bibinfo  {journal} {Nat. Commun.}\ }\textbf {\bibinfo {volume} {13}},\ \bibinfo {pages} {2453} (\bibinfo {year} {2022})}\BibitemShut {NoStop}%
\bibitem [{\citenamefont {Batatia}\ \emph {et~al.}(2022)\citenamefont {Batatia}, \citenamefont {Kovacs}, \citenamefont {Simm}, \citenamefont {Ortner},\ and\ \citenamefont {Csanyi}}]{Batatia2022}%
  \BibitemOpen
  \bibfield  {author} {\bibinfo {author} {\bibfnamefont {I.}~\bibnamefont {Batatia}}, \bibinfo {author} {\bibfnamefont {D.~P.}\ \bibnamefont {Kovacs}}, \bibinfo {author} {\bibfnamefont {G.}~\bibnamefont {Simm}}, \bibinfo {author} {\bibfnamefont {C.}~\bibnamefont {Ortner}}, \ and\ \bibinfo {author} {\bibfnamefont {G.}~\bibnamefont {Csanyi}},\ }in\ \href {https://proceedings.neurips.cc/paper_files/paper/2022/file/4a36c3c51af11ed9f34615b81edb5bbc-Paper-Conference.pdf} {\emph {\bibinfo {booktitle} {Advances in Neural Information Processing Systems}}},\ Vol.~\bibinfo {volume} {35},\ \bibinfo {editor} {edited by\ \bibinfo {editor} {\bibfnamefont {S.}~\bibnamefont {Koyejo}}, \bibinfo {editor} {\bibfnamefont {S.}~\bibnamefont {Mohamed}}, \bibinfo {editor} {\bibfnamefont {A.}~\bibnamefont {Agarwal}}, \bibinfo {editor} {\bibfnamefont {D.}~\bibnamefont {Belgrave}}, \bibinfo {editor} {\bibfnamefont {K.}~\bibnamefont {Cho}}, \ and\ \bibinfo {editor} {\bibfnamefont {A.}~\bibnamefont {Oh}}}\ (\bibinfo  {publisher} {Curran
  Associates, Inc.},\ \bibinfo {year} {2022})\ pp.\ \bibinfo {pages} {11423--11436}\BibitemShut {NoStop}%
\bibitem [{\citenamefont {Frank}, \citenamefont {Unke},\ and\ \citenamefont {M{\"u}ller}(2022)}]{frank2022so3krates}%
  \BibitemOpen
  \bibfield  {author} {\bibinfo {author} {\bibfnamefont {T.}~\bibnamefont {Frank}}, \bibinfo {author} {\bibfnamefont {O.}~\bibnamefont {Unke}}, \ and\ \bibinfo {author} {\bibfnamefont {K.-R.}\ \bibnamefont {M{\"u}ller}},\ }\href@noop {} {\bibfield  {journal} {\bibinfo  {journal} {Advances in Neural Information Processing Systems}\ }\textbf {\bibinfo {volume} {35}},\ \bibinfo {pages} {29400} (\bibinfo {year} {2022})}\BibitemShut {NoStop}%
\bibitem [{\citenamefont {Frank}\ \emph {et~al.}(2024)\citenamefont {Frank}, \citenamefont {Unke}, \citenamefont {M{\"u}ller},\ and\ \citenamefont {Chmiela}}]{frank2024euclidean}%
  \BibitemOpen
  \bibfield  {author} {\bibinfo {author} {\bibfnamefont {T.}~\bibnamefont {Frank}}, \bibinfo {author} {\bibfnamefont {O.}~\bibnamefont {Unke}}, \bibinfo {author} {\bibfnamefont {K.-R.}\ \bibnamefont {M{\"u}ller}}, \ and\ \bibinfo {author} {\bibfnamefont {S.}~\bibnamefont {Chmiela}},\ }\href@noop {} {\bibfield  {journal} {\bibinfo  {journal} {Nat. Commun.}\ }\textbf {\bibinfo {volume} {15}},\ \bibinfo {pages} {6539} (\bibinfo {year} {2024})}\BibitemShut {NoStop}%
\bibitem [{\citenamefont {Kovács}\ \emph {et~al.}(2025)\citenamefont {Kovács}, \citenamefont {Moore}, \citenamefont {Browning}, \citenamefont {Batatia}, \citenamefont {Horton}, \citenamefont {Pu}, \citenamefont {Kapil}, \citenamefont {Witt}, \citenamefont {Magdău}, \citenamefont {Cole},\ and\ \citenamefont {Csányi}}]{Kovcs2025}%
  \BibitemOpen
  \bibfield  {author} {\bibinfo {author} {\bibfnamefont {D.~P.}\ \bibnamefont {Kovács}}, \bibinfo {author} {\bibfnamefont {J.~H.}\ \bibnamefont {Moore}}, \bibinfo {author} {\bibfnamefont {N.~J.}\ \bibnamefont {Browning}}, \bibinfo {author} {\bibfnamefont {I.}~\bibnamefont {Batatia}}, \bibinfo {author} {\bibfnamefont {J.~T.}\ \bibnamefont {Horton}}, \bibinfo {author} {\bibfnamefont {Y.}~\bibnamefont {Pu}}, \bibinfo {author} {\bibfnamefont {V.}~\bibnamefont {Kapil}}, \bibinfo {author} {\bibfnamefont {W.~C.}\ \bibnamefont {Witt}}, \bibinfo {author} {\bibfnamefont {I.-B.}\ \bibnamefont {Magdău}}, \bibinfo {author} {\bibfnamefont {D.~J.}\ \bibnamefont {Cole}}, \ and\ \bibinfo {author} {\bibfnamefont {G.}~\bibnamefont {Csányi}},\ }\href {\doibase 10.1021/jacs.4c07099} {\bibfield  {journal} {\bibinfo  {journal} {J. Am. Chem. Soc.}\ }\textbf {\bibinfo {volume} {147}},\ \bibinfo {pages} {17598–17611} (\bibinfo {year} {2025})}\BibitemShut {NoStop}%
\bibitem [{\citenamefont {Žugec}, \citenamefont {Geilhufe},\ and\ \citenamefont {Lončarić}(2024)}]{Zugec2024}%
  \BibitemOpen
  \bibfield  {author} {\bibinfo {author} {\bibfnamefont {I.}~\bibnamefont {Žugec}}, \bibinfo {author} {\bibfnamefont {R.~M.}\ \bibnamefont {Geilhufe}}, \ and\ \bibinfo {author} {\bibfnamefont {I.}~\bibnamefont {Lončarić}},\ }\href {\doibase 10.1063/5.0196232} {\bibfield  {journal} {\bibinfo  {journal} {J. Chem. Phys.}\ }\textbf {\bibinfo {volume} {160}},\ \bibinfo {pages} {154106} (\bibinfo {year} {2024})}\BibitemShut {NoStop}%
\bibitem [{\citenamefont {Mladineo}\ and\ \citenamefont {Lončarić}(2024)}]{Mladineo2024}%
  \BibitemOpen
  \bibfield  {author} {\bibinfo {author} {\bibfnamefont {B.}~\bibnamefont {Mladineo}}\ and\ \bibinfo {author} {\bibfnamefont {I.}~\bibnamefont {Lončarić}},\ }\href {\doibase 10.1021/acs.cgd.4c00905} {\bibfield  {journal} {\bibinfo  {journal} {Cryst. Growth Des.}\ }\textbf {\bibinfo {volume} {24}},\ \bibinfo {pages} {8167–8173} (\bibinfo {year} {2024})}\BibitemShut {NoStop}%
\bibitem [{\citenamefont {Zhao}\ \emph {et~al.}(2025)\citenamefont {Zhao}, \citenamefont {Liu}, \citenamefont {Qu}, \citenamefont {Cooper},\ and\ \citenamefont {Chen}}]{Zhao2025}%
  \BibitemOpen
  \bibfield  {author} {\bibinfo {author} {\bibfnamefont {C.}~\bibnamefont {Zhao}}, \bibinfo {author} {\bibfnamefont {H.}~\bibnamefont {Liu}}, \bibinfo {author} {\bibfnamefont {D.-H.}\ \bibnamefont {Qu}}, \bibinfo {author} {\bibfnamefont {A.~I.}\ \bibnamefont {Cooper}}, \ and\ \bibinfo {author} {\bibfnamefont {L.}~\bibnamefont {Chen}},\ }\href {\doibase 10.1039/d4sc06467d} {\bibfield  {journal} {\bibinfo  {journal} {Chem. Sci.}\ }\textbf {\bibinfo {volume} {16}},\ \bibinfo {pages} {2363} (\bibinfo {year} {2025})}\BibitemShut {NoStop}%
\bibitem [{\citenamefont {Nickerson}\ and\ \citenamefont {Johnson}(2025)}]{Nickerson2025}%
  \BibitemOpen
  \bibfield  {author} {\bibinfo {author} {\bibfnamefont {C.~J.}\ \bibnamefont {Nickerson}}\ and\ \bibinfo {author} {\bibfnamefont {E.~R.}\ \bibnamefont {Johnson}},\ }\href {\doibase 10.1039/d5cp00593k} {\bibfield  {journal} {\bibinfo  {journal} {Phys. Chem. Chem. Phys.}\ }\textbf {\bibinfo {volume} {27}},\ \bibinfo {pages} {11930–11940} (\bibinfo {year} {2025})}\BibitemShut {NoStop}%
\bibitem [{\citenamefont {Henkelman}, \citenamefont {Uberuaga},\ and\ \citenamefont {J{\'{o}}nsson}(2000)}]{Henkelman2000}%
  \BibitemOpen
  \bibfield  {author} {\bibinfo {author} {\bibfnamefont {G.}~\bibnamefont {Henkelman}}, \bibinfo {author} {\bibfnamefont {B.~P.}\ \bibnamefont {Uberuaga}}, \ and\ \bibinfo {author} {\bibfnamefont {H.}~\bibnamefont {J{\'{o}}nsson}},\ }\href {\doibase 10.1063/1.1329672} {\bibfield  {journal} {\bibinfo  {journal} {J. Chem. Phys.}\ }\textbf {\bibinfo {volume} {113}},\ \bibinfo {pages} {9901} (\bibinfo {year} {2000})}\BibitemShut {NoStop}%
\bibitem [{\citenamefont {Gelli}\ \emph {et~al.}(2026)\citenamefont {Gelli}, \citenamefont {Casassa}, \citenamefont {Rimola},\ and\ \citenamefont {Ribaldone}}]{Gelli2026}%
  \BibitemOpen
  \bibfield  {author} {\bibinfo {author} {\bibfnamefont {A.}~\bibnamefont {Gelli}}, \bibinfo {author} {\bibfnamefont {S.}~\bibnamefont {Casassa}}, \bibinfo {author} {\bibfnamefont {A.}~\bibnamefont {Rimola}}, \ and\ \bibinfo {author} {\bibfnamefont {C.}~\bibnamefont {Ribaldone}},\ }\href {\doibase 10.1063/5.0309524} {\bibfield  {journal} {\bibinfo  {journal} {J. Chem. Phys.}\ }\textbf {\bibinfo {volume} {164}},\ \bibinfo {pages} {044118} (\bibinfo {year} {2026})}\BibitemShut {NoStop}%
\bibitem [{\citenamefont {Rohling}\ \emph {et~al.}(2019)\citenamefont {Rohling}, \citenamefont {Tranca}, \citenamefont {Hensen},\ and\ \citenamefont {Pidko}}]{Rohling2018}%
  \BibitemOpen
  \bibfield  {author} {\bibinfo {author} {\bibfnamefont {R.~Y.}\ \bibnamefont {Rohling}}, \bibinfo {author} {\bibfnamefont {I.~C.}\ \bibnamefont {Tranca}}, \bibinfo {author} {\bibfnamefont {E.~J.~M.}\ \bibnamefont {Hensen}}, \ and\ \bibinfo {author} {\bibfnamefont {E.~A.}\ \bibnamefont {Pidko}},\ }\href {\doibase 10.1021/acscatal.8b03482} {\bibfield  {journal} {\bibinfo  {journal} {ACS Catal.}\ }\textbf {\bibinfo {volume} {9}},\ \bibinfo {pages} {376–391} (\bibinfo {year} {2019})}\BibitemShut {NoStop}%
\bibitem [{\citenamefont {Sheppard}\ \emph {et~al.}(2012)\citenamefont {Sheppard}, \citenamefont {Xiao}, \citenamefont {Chemelewski}, \citenamefont {Johnson},\ and\ \citenamefont {Henkelman}}]{Sheppard2012}%
  \BibitemOpen
  \bibfield  {author} {\bibinfo {author} {\bibfnamefont {D.}~\bibnamefont {Sheppard}}, \bibinfo {author} {\bibfnamefont {P.}~\bibnamefont {Xiao}}, \bibinfo {author} {\bibfnamefont {W.}~\bibnamefont {Chemelewski}}, \bibinfo {author} {\bibfnamefont {D.~D.}\ \bibnamefont {Johnson}}, \ and\ \bibinfo {author} {\bibfnamefont {G.}~\bibnamefont {Henkelman}},\ }\href {\doibase 10.1063/1.3684549} {\bibfield  {journal} {\bibinfo  {journal} {J. Chem. Phys.}\ }\textbf {\bibinfo {volume} {136}},\ \bibinfo {pages} {074103} (\bibinfo {year} {2012})}\BibitemShut {NoStop}%
\bibitem [{\citenamefont {Shi}\ \emph {et~al.}(2022)\citenamefont {Shi}, \citenamefont {Chen}, \citenamefont {Wang}, \citenamefont {Xu}, \citenamefont {Liu},\ and\ \citenamefont {Zhang}}]{Shi2022}%
  \BibitemOpen
  \bibfield  {author} {\bibinfo {author} {\bibfnamefont {Q.}~\bibnamefont {Shi}}, \bibinfo {author} {\bibfnamefont {H.}~\bibnamefont {Chen}}, \bibinfo {author} {\bibfnamefont {Y.}~\bibnamefont {Wang}}, \bibinfo {author} {\bibfnamefont {J.}~\bibnamefont {Xu}}, \bibinfo {author} {\bibfnamefont {Z.}~\bibnamefont {Liu}}, \ and\ \bibinfo {author} {\bibfnamefont {C.}~\bibnamefont {Zhang}},\ }\href {\doibase 10.1016/j.ijpharm.2021.121320} {\bibfield  {journal} {\bibinfo  {journal} {Int. J. Pharm.}\ }\textbf {\bibinfo {volume} {611}},\ \bibinfo {pages} {121320} (\bibinfo {year} {2022})}\BibitemShut {NoStop}%
\bibitem [{\citenamefont {Haleblian}\ and\ \citenamefont {McCrone}(1969)}]{Haleblian1969}%
  \BibitemOpen
  \bibfield  {author} {\bibinfo {author} {\bibfnamefont {J.}~\bibnamefont {Haleblian}}\ and\ \bibinfo {author} {\bibfnamefont {W.}~\bibnamefont {McCrone}},\ }\href {\doibase 10.1002/jps.2600580802} {\bibfield  {journal} {\bibinfo  {journal} {J. Pharm. Sci}\ }\textbf {\bibinfo {volume} {58}},\ \bibinfo {pages} {911–929} (\bibinfo {year} {1969})}\BibitemShut {NoStop}%
\bibitem [{\citenamefont {Lee}, \citenamefont {Erdemir},\ and\ \citenamefont {Myerson}(2011)}]{Lee2011}%
  \BibitemOpen
  \bibfield  {author} {\bibinfo {author} {\bibfnamefont {A.~Y.}\ \bibnamefont {Lee}}, \bibinfo {author} {\bibfnamefont {D.}~\bibnamefont {Erdemir}}, \ and\ \bibinfo {author} {\bibfnamefont {A.~S.}\ \bibnamefont {Myerson}},\ }\href {\doibase 10.1146/annurev-chembioeng-061010-114224} {\bibfield  {journal} {\bibinfo  {journal} {Annual Review of Chemical and Biomolecular Engineering}\ }\textbf {\bibinfo {volume} {2}},\ \bibinfo {pages} {259–280} (\bibinfo {year} {2011})}\BibitemShut {NoStop}%
\bibitem [{\citenamefont {Braga}, \citenamefont {Casali},\ and\ \citenamefont {Grepioni}(2022)}]{Braga2022}%
  \BibitemOpen
  \bibfield  {author} {\bibinfo {author} {\bibfnamefont {D.}~\bibnamefont {Braga}}, \bibinfo {author} {\bibfnamefont {L.}~\bibnamefont {Casali}}, \ and\ \bibinfo {author} {\bibfnamefont {F.}~\bibnamefont {Grepioni}},\ }\href {\doibase 10.3390/ijms23169013} {\bibfield  {journal} {\bibinfo  {journal} {Int. J. Mol. Sci.}\ }\textbf {\bibinfo {volume} {23}},\ \bibinfo {pages} {9013} (\bibinfo {year} {2022})}\BibitemShut {NoStop}%
\bibitem [{\citenamefont {Hunnisett}\ \emph {et~al.}(2024{\natexlab{a}})\citenamefont {Hunnisett}, \citenamefont {Nyman}, \citenamefont {Francia}, \citenamefont {Abraham}, \citenamefont {Adjiman}, \citenamefont {Aitipamula}, \citenamefont {Alkhidir}, \citenamefont {Almehairbi}, \citenamefont {Anelli}, \citenamefont {Anstine}, \citenamefont {Anthony}, \citenamefont {Arnold}, \citenamefont {Bahrami}, \citenamefont {Bellucci}, \citenamefont {Bhardwaj}, \citenamefont {Bier}, \citenamefont {Bis}, \citenamefont {Boese}, \citenamefont {Bowskill}, \citenamefont {Bramley}, \citenamefont {Brandenburg}, \citenamefont {Braun}, \citenamefont {Butler}, \citenamefont {Cadden}, \citenamefont {Carino}, \citenamefont {Chan}, \citenamefont {Chang}, \citenamefont {Cheng}, \citenamefont {Clarke}, \citenamefont {Coles}, \citenamefont {Cooper}, \citenamefont {Couch}, \citenamefont {Cuadrado}, \citenamefont {Darden}, \citenamefont {Day}, \citenamefont {Dietrich}, \citenamefont {Ding}, \citenamefont {DiPasquale}, \citenamefont
  {Dhokale}, \citenamefont {van Eijck}, \citenamefont {Elsegood}, \citenamefont {Firaha}, \citenamefont {Fu}, \citenamefont {Fukuzawa}, \citenamefont {Glover}, \citenamefont {Goto}, \citenamefont {Greenwell}, \citenamefont {Guo}, \citenamefont {Harter}, \citenamefont {Helfferich}, \citenamefont {Hofmann}, \citenamefont {Hoja}, \citenamefont {Hone}, \citenamefont {Hong}, \citenamefont {Hutchison}, \citenamefont {Ikabata}, \citenamefont {Isayev}, \citenamefont {Ishaque}, \citenamefont {Jain}, \citenamefont {Jin}, \citenamefont {Jing}, \citenamefont {Johnson}, \citenamefont {Jones}, \citenamefont {Jose}, \citenamefont {Kabova}, \citenamefont {Keates}, \citenamefont {Kelly}, \citenamefont {Khakimov}, \citenamefont {Konstantinopoulos}, \citenamefont {Kuleshova}, \citenamefont {Li}, \citenamefont {Lin}, \citenamefont {List}, \citenamefont {Liu}, \citenamefont {Liu}, \citenamefont {Liu}, \citenamefont {Liu}, \citenamefont {Lubach}, \citenamefont {Marom}, \citenamefont {Maryewski}, \citenamefont {Matsui},
  \citenamefont {Mattei}, \citenamefont {Mayo}, \citenamefont {Melkumov}, \citenamefont {Mohamed}, \citenamefont {Momenzadeh~Abardeh}, \citenamefont {Muddana}, \citenamefont {Nakayama}, \citenamefont {Nayal}, \citenamefont {Neumann}, \citenamefont {Nikhar}, \citenamefont {Obata}, \citenamefont {O’Connor}, \citenamefont {Oganov}, \citenamefont {Okuwaki}, \citenamefont {Otero-de-la Roza}, \citenamefont {Pantelides}, \citenamefont {Parkin}, \citenamefont {Pickard}, \citenamefont {Pilia}, \citenamefont {Pivina}, \citenamefont {Podeszwa}, \citenamefont {Price}, \citenamefont {Price}, \citenamefont {Price}, \citenamefont {Probert}, \citenamefont {Pulido}, \citenamefont {Ramteke}, \citenamefont {Rehman}, \citenamefont {Reutzel-Edens}, \citenamefont {Rogal}, \citenamefont {Ross}, \citenamefont {Rumson}, \citenamefont {Sadiq}, \citenamefont {Saeed}, \citenamefont {Salimi}, \citenamefont {Salvalaglio}, \citenamefont {Sanders~de Almada}, \citenamefont {Sasikumar}, \citenamefont {Sekharan}, \citenamefont {Shang},
  \citenamefont {Shankland}, \citenamefont {Shinohara}, \citenamefont {Shi}, \citenamefont {Shi}, \citenamefont {Skillman}, \citenamefont {Song}, \citenamefont {Strasser}, \citenamefont {van~de Streek}, \citenamefont {Sugden}, \citenamefont {Sun}, \citenamefont {Szalewicz}, \citenamefont {Tan}, \citenamefont {Tan}, \citenamefont {Tarczynski}, \citenamefont {Taylor}, \citenamefont {Tkatchenko}, \citenamefont {Tom}, \citenamefont {Tuckerman}, \citenamefont {Utsumi}, \citenamefont {Vogt-Maranto}, \citenamefont {Weatherston}, \citenamefont {Wilkinson}, \citenamefont {Willacy}, \citenamefont {Wojtas}, \citenamefont {Woollam}, \citenamefont {Yang}, \citenamefont {Yonemochi}, \citenamefont {Yue}, \citenamefont {Zeng}, \citenamefont {Zhang}, \citenamefont {Zhou}, \citenamefont {Zhou}, \citenamefont {Zubatyuk},\ and\ \citenamefont {Cole}}]{Hunnisett2024}%
  \BibitemOpen
  \bibfield  {author} {\bibinfo {author} {\bibfnamefont {L.~M.}\ \bibnamefont {Hunnisett}}, \bibinfo {author} {\bibfnamefont {J.}~\bibnamefont {Nyman}}, \bibinfo {author} {\bibfnamefont {N.}~\bibnamefont {Francia}}, \bibinfo {author} {\bibfnamefont {N.~S.}\ \bibnamefont {Abraham}}, \bibinfo {author} {\bibfnamefont {C.~S.}\ \bibnamefont {Adjiman}}, \bibinfo {author} {\bibfnamefont {S.}~\bibnamefont {Aitipamula}}, \bibinfo {author} {\bibfnamefont {T.}~\bibnamefont {Alkhidir}}, \bibinfo {author} {\bibfnamefont {M.}~\bibnamefont {Almehairbi}}, \bibinfo {author} {\bibfnamefont {A.}~\bibnamefont {Anelli}}, \bibinfo {author} {\bibfnamefont {D.~M.}\ \bibnamefont {Anstine}}, \bibinfo {author} {\bibfnamefont {J.~E.}\ \bibnamefont {Anthony}}, \bibinfo {author} {\bibfnamefont {J.~E.}\ \bibnamefont {Arnold}}, \bibinfo {author} {\bibfnamefont {F.}~\bibnamefont {Bahrami}}, \bibinfo {author} {\bibfnamefont {M.~A.}\ \bibnamefont {Bellucci}}, \bibinfo {author} {\bibfnamefont {R.~M.}\ \bibnamefont {Bhardwaj}}, \bibinfo {author}
  {\bibfnamefont {I.}~\bibnamefont {Bier}}, \bibinfo {author} {\bibfnamefont {J.~A.}\ \bibnamefont {Bis}}, \bibinfo {author} {\bibfnamefont {A.~D.}\ \bibnamefont {Boese}}, \bibinfo {author} {\bibfnamefont {D.~H.}\ \bibnamefont {Bowskill}}, \bibinfo {author} {\bibfnamefont {J.}~\bibnamefont {Bramley}}, \bibinfo {author} {\bibfnamefont {J.~G.}\ \bibnamefont {Brandenburg}}, \bibinfo {author} {\bibfnamefont {D.~E.}\ \bibnamefont {Braun}}, \bibinfo {author} {\bibfnamefont {P.~W.~V.}\ \bibnamefont {Butler}}, \bibinfo {author} {\bibfnamefont {J.}~\bibnamefont {Cadden}}, \bibinfo {author} {\bibfnamefont {S.}~\bibnamefont {Carino}}, \bibinfo {author} {\bibfnamefont {E.~J.}\ \bibnamefont {Chan}}, \bibinfo {author} {\bibfnamefont {C.}~\bibnamefont {Chang}}, \bibinfo {author} {\bibfnamefont {B.}~\bibnamefont {Cheng}}, \bibinfo {author} {\bibfnamefont {S.~M.}\ \bibnamefont {Clarke}}, \bibinfo {author} {\bibfnamefont {S.~J.}\ \bibnamefont {Coles}}, \bibinfo {author} {\bibfnamefont {R.~I.}\ \bibnamefont {Cooper}}, \bibinfo
  {author} {\bibfnamefont {R.}~\bibnamefont {Couch}}, \bibinfo {author} {\bibfnamefont {R.}~\bibnamefont {Cuadrado}}, \bibinfo {author} {\bibfnamefont {T.}~\bibnamefont {Darden}}, \bibinfo {author} {\bibfnamefont {G.~M.}\ \bibnamefont {Day}}, \bibinfo {author} {\bibfnamefont {H.}~\bibnamefont {Dietrich}}, \bibinfo {author} {\bibfnamefont {Y.}~\bibnamefont {Ding}}, \bibinfo {author} {\bibfnamefont {A.}~\bibnamefont {DiPasquale}}, \bibinfo {author} {\bibfnamefont {B.}~\bibnamefont {Dhokale}}, \bibinfo {author} {\bibfnamefont {B.~P.}\ \bibnamefont {van Eijck}}, \bibinfo {author} {\bibfnamefont {M.~R.~J.}\ \bibnamefont {Elsegood}}, \bibinfo {author} {\bibfnamefont {D.}~\bibnamefont {Firaha}}, \bibinfo {author} {\bibfnamefont {W.}~\bibnamefont {Fu}}, \bibinfo {author} {\bibfnamefont {K.}~\bibnamefont {Fukuzawa}}, \bibinfo {author} {\bibfnamefont {J.}~\bibnamefont {Glover}}, \bibinfo {author} {\bibfnamefont {H.}~\bibnamefont {Goto}}, \bibinfo {author} {\bibfnamefont {C.}~\bibnamefont {Greenwell}}, \bibinfo {author}
  {\bibfnamefont {R.}~\bibnamefont {Guo}}, \bibinfo {author} {\bibfnamefont {J.}~\bibnamefont {Harter}}, \bibinfo {author} {\bibfnamefont {J.}~\bibnamefont {Helfferich}}, \bibinfo {author} {\bibfnamefont {D.~W.~M.}\ \bibnamefont {Hofmann}}, \bibinfo {author} {\bibfnamefont {J.}~\bibnamefont {Hoja}}, \bibinfo {author} {\bibfnamefont {J.}~\bibnamefont {Hone}}, \bibinfo {author} {\bibfnamefont {R.}~\bibnamefont {Hong}}, \bibinfo {author} {\bibfnamefont {G.}~\bibnamefont {Hutchison}}, \bibinfo {author} {\bibfnamefont {Y.}~\bibnamefont {Ikabata}}, \bibinfo {author} {\bibfnamefont {O.}~\bibnamefont {Isayev}}, \bibinfo {author} {\bibfnamefont {O.}~\bibnamefont {Ishaque}}, \bibinfo {author} {\bibfnamefont {V.}~\bibnamefont {Jain}}, \bibinfo {author} {\bibfnamefont {Y.}~\bibnamefont {Jin}}, \bibinfo {author} {\bibfnamefont {A.}~\bibnamefont {Jing}}, \bibinfo {author} {\bibfnamefont {E.~R.}\ \bibnamefont {Johnson}}, \bibinfo {author} {\bibfnamefont {I.}~\bibnamefont {Jones}}, \bibinfo {author} {\bibfnamefont
  {K.~V.~J.}\ \bibnamefont {Jose}}, \bibinfo {author} {\bibfnamefont {E.~A.}\ \bibnamefont {Kabova}}, \bibinfo {author} {\bibfnamefont {A.}~\bibnamefont {Keates}}, \bibinfo {author} {\bibfnamefont {P.~F.}\ \bibnamefont {Kelly}}, \bibinfo {author} {\bibfnamefont {D.}~\bibnamefont {Khakimov}}, \bibinfo {author} {\bibfnamefont {S.}~\bibnamefont {Konstantinopoulos}}, \bibinfo {author} {\bibfnamefont {L.~N.}\ \bibnamefont {Kuleshova}}, \bibinfo {author} {\bibfnamefont {H.}~\bibnamefont {Li}}, \bibinfo {author} {\bibfnamefont {X.}~\bibnamefont {Lin}}, \bibinfo {author} {\bibfnamefont {A.}~\bibnamefont {List}}, \bibinfo {author} {\bibfnamefont {C.}~\bibnamefont {Liu}}, \bibinfo {author} {\bibfnamefont {Y.~M.}\ \bibnamefont {Liu}}, \bibinfo {author} {\bibfnamefont {Z.}~\bibnamefont {Liu}}, \bibinfo {author} {\bibfnamefont {Z.-P.}\ \bibnamefont {Liu}}, \bibinfo {author} {\bibfnamefont {J.~W.}\ \bibnamefont {Lubach}}, \bibinfo {author} {\bibfnamefont {N.}~\bibnamefont {Marom}}, \bibinfo {author} {\bibfnamefont {A.~A.}\
  \bibnamefont {Maryewski}}, \bibinfo {author} {\bibfnamefont {H.}~\bibnamefont {Matsui}}, \bibinfo {author} {\bibfnamefont {A.}~\bibnamefont {Mattei}}, \bibinfo {author} {\bibfnamefont {R.~A.}\ \bibnamefont {Mayo}}, \bibinfo {author} {\bibfnamefont {J.~W.}\ \bibnamefont {Melkumov}}, \bibinfo {author} {\bibfnamefont {S.}~\bibnamefont {Mohamed}}, \bibinfo {author} {\bibfnamefont {Z.}~\bibnamefont {Momenzadeh~Abardeh}}, \bibinfo {author} {\bibfnamefont {H.~S.}\ \bibnamefont {Muddana}}, \bibinfo {author} {\bibfnamefont {N.}~\bibnamefont {Nakayama}}, \bibinfo {author} {\bibfnamefont {K.~S.}\ \bibnamefont {Nayal}}, \bibinfo {author} {\bibfnamefont {M.~A.}\ \bibnamefont {Neumann}}, \bibinfo {author} {\bibfnamefont {R.}~\bibnamefont {Nikhar}}, \bibinfo {author} {\bibfnamefont {S.}~\bibnamefont {Obata}}, \bibinfo {author} {\bibfnamefont {D.}~\bibnamefont {O’Connor}}, \bibinfo {author} {\bibfnamefont {A.~R.}\ \bibnamefont {Oganov}}, \bibinfo {author} {\bibfnamefont {K.}~\bibnamefont {Okuwaki}}, \bibinfo {author}
  {\bibfnamefont {A.}~\bibnamefont {Otero-de-la Roza}}, \bibinfo {author} {\bibfnamefont {C.~C.}\ \bibnamefont {Pantelides}}, \bibinfo {author} {\bibfnamefont {S.}~\bibnamefont {Parkin}}, \bibinfo {author} {\bibfnamefont {C.~J.}\ \bibnamefont {Pickard}}, \bibinfo {author} {\bibfnamefont {L.}~\bibnamefont {Pilia}}, \bibinfo {author} {\bibfnamefont {T.}~\bibnamefont {Pivina}}, \bibinfo {author} {\bibfnamefont {R.}~\bibnamefont {Podeszwa}}, \bibinfo {author} {\bibfnamefont {A.~J.~A.}\ \bibnamefont {Price}}, \bibinfo {author} {\bibfnamefont {L.~S.}\ \bibnamefont {Price}}, \bibinfo {author} {\bibfnamefont {S.~L.}\ \bibnamefont {Price}}, \bibinfo {author} {\bibfnamefont {M.~R.}\ \bibnamefont {Probert}}, \bibinfo {author} {\bibfnamefont {A.}~\bibnamefont {Pulido}}, \bibinfo {author} {\bibfnamefont {G.~R.}\ \bibnamefont {Ramteke}}, \bibinfo {author} {\bibfnamefont {A.~U.}\ \bibnamefont {Rehman}}, \bibinfo {author} {\bibfnamefont {S.~M.}\ \bibnamefont {Reutzel-Edens}}, \bibinfo {author} {\bibfnamefont
  {J.}~\bibnamefont {Rogal}}, \bibinfo {author} {\bibfnamefont {M.~J.}\ \bibnamefont {Ross}}, \bibinfo {author} {\bibfnamefont {A.~F.}\ \bibnamefont {Rumson}}, \bibinfo {author} {\bibfnamefont {G.}~\bibnamefont {Sadiq}}, \bibinfo {author} {\bibfnamefont {Z.~M.}\ \bibnamefont {Saeed}}, \bibinfo {author} {\bibfnamefont {A.}~\bibnamefont {Salimi}}, \bibinfo {author} {\bibfnamefont {M.}~\bibnamefont {Salvalaglio}}, \bibinfo {author} {\bibfnamefont {L.}~\bibnamefont {Sanders~de Almada}}, \bibinfo {author} {\bibfnamefont {K.}~\bibnamefont {Sasikumar}}, \bibinfo {author} {\bibfnamefont {S.}~\bibnamefont {Sekharan}}, \bibinfo {author} {\bibfnamefont {C.}~\bibnamefont {Shang}}, \bibinfo {author} {\bibfnamefont {K.}~\bibnamefont {Shankland}}, \bibinfo {author} {\bibfnamefont {K.}~\bibnamefont {Shinohara}}, \bibinfo {author} {\bibfnamefont {B.}~\bibnamefont {Shi}}, \bibinfo {author} {\bibfnamefont {X.}~\bibnamefont {Shi}}, \bibinfo {author} {\bibfnamefont {A.~G.}\ \bibnamefont {Skillman}}, \bibinfo {author}
  {\bibfnamefont {H.}~\bibnamefont {Song}}, \bibinfo {author} {\bibfnamefont {N.}~\bibnamefont {Strasser}}, \bibinfo {author} {\bibfnamefont {J.}~\bibnamefont {van~de Streek}}, \bibinfo {author} {\bibfnamefont {I.~J.}\ \bibnamefont {Sugden}}, \bibinfo {author} {\bibfnamefont {G.}~\bibnamefont {Sun}}, \bibinfo {author} {\bibfnamefont {K.}~\bibnamefont {Szalewicz}}, \bibinfo {author} {\bibfnamefont {B.~I.}\ \bibnamefont {Tan}}, \bibinfo {author} {\bibfnamefont {L.}~\bibnamefont {Tan}}, \bibinfo {author} {\bibfnamefont {F.}~\bibnamefont {Tarczynski}}, \bibinfo {author} {\bibfnamefont {C.~R.}\ \bibnamefont {Taylor}}, \bibinfo {author} {\bibfnamefont {A.}~\bibnamefont {Tkatchenko}}, \bibinfo {author} {\bibfnamefont {R.}~\bibnamefont {Tom}}, \bibinfo {author} {\bibfnamefont {M.~E.}\ \bibnamefont {Tuckerman}}, \bibinfo {author} {\bibfnamefont {Y.}~\bibnamefont {Utsumi}}, \bibinfo {author} {\bibfnamefont {L.}~\bibnamefont {Vogt-Maranto}}, \bibinfo {author} {\bibfnamefont {J.}~\bibnamefont {Weatherston}}, \bibinfo
  {author} {\bibfnamefont {L.~J.}\ \bibnamefont {Wilkinson}}, \bibinfo {author} {\bibfnamefont {R.~D.}\ \bibnamefont {Willacy}}, \bibinfo {author} {\bibfnamefont {L.}~\bibnamefont {Wojtas}}, \bibinfo {author} {\bibfnamefont {G.~R.}\ \bibnamefont {Woollam}}, \bibinfo {author} {\bibfnamefont {Z.}~\bibnamefont {Yang}}, \bibinfo {author} {\bibfnamefont {E.}~\bibnamefont {Yonemochi}}, \bibinfo {author} {\bibfnamefont {X.}~\bibnamefont {Yue}}, \bibinfo {author} {\bibfnamefont {Q.}~\bibnamefont {Zeng}}, \bibinfo {author} {\bibfnamefont {Y.}~\bibnamefont {Zhang}}, \bibinfo {author} {\bibfnamefont {T.}~\bibnamefont {Zhou}}, \bibinfo {author} {\bibfnamefont {Y.}~\bibnamefont {Zhou}}, \bibinfo {author} {\bibfnamefont {R.}~\bibnamefont {Zubatyuk}}, \ and\ \bibinfo {author} {\bibfnamefont {J.~C.}\ \bibnamefont {Cole}},\ }\href {\doibase 10.1107/s2052520624007492} {\bibfield  {journal} {\bibinfo  {journal} {Acta Crystallogr. Sect. B: Struct. Sci.}\ }\textbf {\bibinfo {volume} {80}},\ \bibinfo {pages} {517–547} (\bibinfo
  {year} {2024}{\natexlab{a}})}\BibitemShut {NoStop}%
\bibitem [{\citenamefont {Hunnisett}\ \emph {et~al.}(2024{\natexlab{b}})\citenamefont {Hunnisett}, \citenamefont {Francia}, \citenamefont {Nyman}, \citenamefont {Abraham}, \citenamefont {Aitipamula}, \citenamefont {Alkhidir}, \citenamefont {Almehairbi}, \citenamefont {Anelli}, \citenamefont {Anstine}, \citenamefont {Anthony}, \citenamefont {Arnold}, \citenamefont {Bahrami}, \citenamefont {Bellucci}, \citenamefont {Beran}, \citenamefont {Bhardwaj}, \citenamefont {Bianco}, \citenamefont {Bis}, \citenamefont {Boese}, \citenamefont {Bramley}, \citenamefont {Braun}, \citenamefont {Butler}, \citenamefont {Cadden}, \citenamefont {Carino}, \citenamefont {Červinka}, \citenamefont {Chan}, \citenamefont {Chang}, \citenamefont {Clarke}, \citenamefont {Coles}, \citenamefont {Cook}, \citenamefont {Cooper}, \citenamefont {Darden}, \citenamefont {Day}, \citenamefont {Deng}, \citenamefont {Dietrich}, \citenamefont {DiPasquale}, \citenamefont {Dhokale}, \citenamefont {van Eijck}, \citenamefont {Elsegood}, \citenamefont
  {Firaha}, \citenamefont {Fu}, \citenamefont {Fukuzawa}, \citenamefont {Galanakis}, \citenamefont {Goto}, \citenamefont {Greenwell}, \citenamefont {Guo}, \citenamefont {Harter}, \citenamefont {Helfferich}, \citenamefont {Hoja}, \citenamefont {Hone}, \citenamefont {Hong}, \citenamefont {Hušák}, \citenamefont {Ikabata}, \citenamefont {Isayev}, \citenamefont {Ishaque}, \citenamefont {Jain}, \citenamefont {Jin}, \citenamefont {Jing}, \citenamefont {Johnson}, \citenamefont {Jones}, \citenamefont {Jose}, \citenamefont {Kabova}, \citenamefont {Keates}, \citenamefont {Kelly}, \citenamefont {Klimeš}, \citenamefont {Kostková}, \citenamefont {Li}, \citenamefont {Lin}, \citenamefont {List}, \citenamefont {Liu}, \citenamefont {Liu}, \citenamefont {Liu}, \citenamefont {Lončarić}, \citenamefont {Lubach}, \citenamefont {Ludík}, \citenamefont {Marom}, \citenamefont {Matsui}, \citenamefont {Mattei}, \citenamefont {Mayo}, \citenamefont {Melkumov}, \citenamefont {Mladineo}, \citenamefont {Mohamed}, \citenamefont
  {Momenzadeh~Abardeh}, \citenamefont {Muddana}, \citenamefont {Nakayama}, \citenamefont {Nayal}, \citenamefont {Neumann}, \citenamefont {Nikhar}, \citenamefont {Obata}, \citenamefont {O’Connor}, \citenamefont {Oganov}, \citenamefont {Okuwaki}, \citenamefont {Otero-de-la Roza}, \citenamefont {Parkin}, \citenamefont {Parunov}, \citenamefont {Podeszwa}, \citenamefont {Price}, \citenamefont {Price}, \citenamefont {Price}, \citenamefont {Probert}, \citenamefont {Pulido}, \citenamefont {Ramteke}, \citenamefont {Rehman}, \citenamefont {Reutzel-Edens}, \citenamefont {Rogal}, \citenamefont {Ross}, \citenamefont {Rumson}, \citenamefont {Sadiq}, \citenamefont {Saeed}, \citenamefont {Salimi}, \citenamefont {Sasikumar}, \citenamefont {Sekharan}, \citenamefont {Shankland}, \citenamefont {Shi}, \citenamefont {Shi}, \citenamefont {Shinohara}, \citenamefont {Skillman}, \citenamefont {Song}, \citenamefont {Strasser}, \citenamefont {van~de Streek}, \citenamefont {Sugden}, \citenamefont {Sun}, \citenamefont {Szalewicz},
  \citenamefont {Tan}, \citenamefont {Tang}, \citenamefont {Tarczynski}, \citenamefont {Taylor}, \citenamefont {Tkatchenko}, \citenamefont {Tom}, \citenamefont {Touš}, \citenamefont {Tuckerman}, \citenamefont {Unzueta}, \citenamefont {Utsumi}, \citenamefont {Vogt-Maranto}, \citenamefont {Weatherston}, \citenamefont {Wilkinson}, \citenamefont {Willacy}, \citenamefont {Wojtas}, \citenamefont {Woollam}, \citenamefont {Yang}, \citenamefont {Yang}, \citenamefont {Yonemochi}, \citenamefont {Yue}, \citenamefont {Zeng}, \citenamefont {Zhou}, \citenamefont {Zhou}, \citenamefont {Zubatyuk},\ and\ \citenamefont {Cole}}]{Hunnisett2024b}%
  \BibitemOpen
  \bibfield  {author} {\bibinfo {author} {\bibfnamefont {L.~M.}\ \bibnamefont {Hunnisett}}, \bibinfo {author} {\bibfnamefont {N.}~\bibnamefont {Francia}}, \bibinfo {author} {\bibfnamefont {J.}~\bibnamefont {Nyman}}, \bibinfo {author} {\bibfnamefont {N.~S.}\ \bibnamefont {Abraham}}, \bibinfo {author} {\bibfnamefont {S.}~\bibnamefont {Aitipamula}}, \bibinfo {author} {\bibfnamefont {T.}~\bibnamefont {Alkhidir}}, \bibinfo {author} {\bibfnamefont {M.}~\bibnamefont {Almehairbi}}, \bibinfo {author} {\bibfnamefont {A.}~\bibnamefont {Anelli}}, \bibinfo {author} {\bibfnamefont {D.~M.}\ \bibnamefont {Anstine}}, \bibinfo {author} {\bibfnamefont {J.~E.}\ \bibnamefont {Anthony}}, \bibinfo {author} {\bibfnamefont {J.~E.}\ \bibnamefont {Arnold}}, \bibinfo {author} {\bibfnamefont {F.}~\bibnamefont {Bahrami}}, \bibinfo {author} {\bibfnamefont {M.~A.}\ \bibnamefont {Bellucci}}, \bibinfo {author} {\bibfnamefont {G.~J.~O.}\ \bibnamefont {Beran}}, \bibinfo {author} {\bibfnamefont {R.~M.}\ \bibnamefont {Bhardwaj}}, \bibinfo
  {author} {\bibfnamefont {R.}~\bibnamefont {Bianco}}, \bibinfo {author} {\bibfnamefont {J.~A.}\ \bibnamefont {Bis}}, \bibinfo {author} {\bibfnamefont {A.~D.}\ \bibnamefont {Boese}}, \bibinfo {author} {\bibfnamefont {J.}~\bibnamefont {Bramley}}, \bibinfo {author} {\bibfnamefont {D.~E.}\ \bibnamefont {Braun}}, \bibinfo {author} {\bibfnamefont {P.~W.~V.}\ \bibnamefont {Butler}}, \bibinfo {author} {\bibfnamefont {J.}~\bibnamefont {Cadden}}, \bibinfo {author} {\bibfnamefont {S.}~\bibnamefont {Carino}}, \bibinfo {author} {\bibfnamefont {C.}~\bibnamefont {Červinka}}, \bibinfo {author} {\bibfnamefont {E.~J.}\ \bibnamefont {Chan}}, \bibinfo {author} {\bibfnamefont {C.}~\bibnamefont {Chang}}, \bibinfo {author} {\bibfnamefont {S.~M.}\ \bibnamefont {Clarke}}, \bibinfo {author} {\bibfnamefont {S.~J.}\ \bibnamefont {Coles}}, \bibinfo {author} {\bibfnamefont {C.~J.}\ \bibnamefont {Cook}}, \bibinfo {author} {\bibfnamefont {R.~I.}\ \bibnamefont {Cooper}}, \bibinfo {author} {\bibfnamefont {T.}~\bibnamefont {Darden}},
  \bibinfo {author} {\bibfnamefont {G.~M.}\ \bibnamefont {Day}}, \bibinfo {author} {\bibfnamefont {W.}~\bibnamefont {Deng}}, \bibinfo {author} {\bibfnamefont {H.}~\bibnamefont {Dietrich}}, \bibinfo {author} {\bibfnamefont {A.}~\bibnamefont {DiPasquale}}, \bibinfo {author} {\bibfnamefont {B.}~\bibnamefont {Dhokale}}, \bibinfo {author} {\bibfnamefont {B.~P.}\ \bibnamefont {van Eijck}}, \bibinfo {author} {\bibfnamefont {M.~R.~J.}\ \bibnamefont {Elsegood}}, \bibinfo {author} {\bibfnamefont {D.}~\bibnamefont {Firaha}}, \bibinfo {author} {\bibfnamefont {W.}~\bibnamefont {Fu}}, \bibinfo {author} {\bibfnamefont {K.}~\bibnamefont {Fukuzawa}}, \bibinfo {author} {\bibfnamefont {N.}~\bibnamefont {Galanakis}}, \bibinfo {author} {\bibfnamefont {H.}~\bibnamefont {Goto}}, \bibinfo {author} {\bibfnamefont {C.}~\bibnamefont {Greenwell}}, \bibinfo {author} {\bibfnamefont {R.}~\bibnamefont {Guo}}, \bibinfo {author} {\bibfnamefont {J.}~\bibnamefont {Harter}}, \bibinfo {author} {\bibfnamefont {J.}~\bibnamefont {Helfferich}},
  \bibinfo {author} {\bibfnamefont {J.}~\bibnamefont {Hoja}}, \bibinfo {author} {\bibfnamefont {J.}~\bibnamefont {Hone}}, \bibinfo {author} {\bibfnamefont {R.}~\bibnamefont {Hong}}, \bibinfo {author} {\bibfnamefont {M.}~\bibnamefont {Hušák}}, \bibinfo {author} {\bibfnamefont {Y.}~\bibnamefont {Ikabata}}, \bibinfo {author} {\bibfnamefont {O.}~\bibnamefont {Isayev}}, \bibinfo {author} {\bibfnamefont {O.}~\bibnamefont {Ishaque}}, \bibinfo {author} {\bibfnamefont {V.}~\bibnamefont {Jain}}, \bibinfo {author} {\bibfnamefont {Y.}~\bibnamefont {Jin}}, \bibinfo {author} {\bibfnamefont {A.}~\bibnamefont {Jing}}, \bibinfo {author} {\bibfnamefont {E.~R.}\ \bibnamefont {Johnson}}, \bibinfo {author} {\bibfnamefont {I.}~\bibnamefont {Jones}}, \bibinfo {author} {\bibfnamefont {K.~V.~J.}\ \bibnamefont {Jose}}, \bibinfo {author} {\bibfnamefont {E.~A.}\ \bibnamefont {Kabova}}, \bibinfo {author} {\bibfnamefont {A.}~\bibnamefont {Keates}}, \bibinfo {author} {\bibfnamefont {P.~F.}\ \bibnamefont {Kelly}}, \bibinfo {author}
  {\bibfnamefont {J.}~\bibnamefont {Klimeš}}, \bibinfo {author} {\bibfnamefont {V.}~\bibnamefont {Kostková}}, \bibinfo {author} {\bibfnamefont {H.}~\bibnamefont {Li}}, \bibinfo {author} {\bibfnamefont {X.}~\bibnamefont {Lin}}, \bibinfo {author} {\bibfnamefont {A.}~\bibnamefont {List}}, \bibinfo {author} {\bibfnamefont {C.}~\bibnamefont {Liu}}, \bibinfo {author} {\bibfnamefont {Y.~M.}\ \bibnamefont {Liu}}, \bibinfo {author} {\bibfnamefont {Z.}~\bibnamefont {Liu}}, \bibinfo {author} {\bibfnamefont {I.}~\bibnamefont {Lončarić}}, \bibinfo {author} {\bibfnamefont {J.~W.}\ \bibnamefont {Lubach}}, \bibinfo {author} {\bibfnamefont {J.}~\bibnamefont {Ludík}}, \bibinfo {author} {\bibfnamefont {N.}~\bibnamefont {Marom}}, \bibinfo {author} {\bibfnamefont {H.}~\bibnamefont {Matsui}}, \bibinfo {author} {\bibfnamefont {A.}~\bibnamefont {Mattei}}, \bibinfo {author} {\bibfnamefont {R.~A.}\ \bibnamefont {Mayo}}, \bibinfo {author} {\bibfnamefont {J.~W.}\ \bibnamefont {Melkumov}}, \bibinfo {author} {\bibfnamefont
  {B.}~\bibnamefont {Mladineo}}, \bibinfo {author} {\bibfnamefont {S.}~\bibnamefont {Mohamed}}, \bibinfo {author} {\bibfnamefont {Z.}~\bibnamefont {Momenzadeh~Abardeh}}, \bibinfo {author} {\bibfnamefont {H.~S.}\ \bibnamefont {Muddana}}, \bibinfo {author} {\bibfnamefont {N.}~\bibnamefont {Nakayama}}, \bibinfo {author} {\bibfnamefont {K.~S.}\ \bibnamefont {Nayal}}, \bibinfo {author} {\bibfnamefont {M.~A.}\ \bibnamefont {Neumann}}, \bibinfo {author} {\bibfnamefont {R.}~\bibnamefont {Nikhar}}, \bibinfo {author} {\bibfnamefont {S.}~\bibnamefont {Obata}}, \bibinfo {author} {\bibfnamefont {D.}~\bibnamefont {O’Connor}}, \bibinfo {author} {\bibfnamefont {A.~R.}\ \bibnamefont {Oganov}}, \bibinfo {author} {\bibfnamefont {K.}~\bibnamefont {Okuwaki}}, \bibinfo {author} {\bibfnamefont {A.}~\bibnamefont {Otero-de-la Roza}}, \bibinfo {author} {\bibfnamefont {S.}~\bibnamefont {Parkin}}, \bibinfo {author} {\bibfnamefont {A.}~\bibnamefont {Parunov}}, \bibinfo {author} {\bibfnamefont {R.}~\bibnamefont {Podeszwa}}, \bibinfo
  {author} {\bibfnamefont {A.~J.~A.}\ \bibnamefont {Price}}, \bibinfo {author} {\bibfnamefont {L.~S.}\ \bibnamefont {Price}}, \bibinfo {author} {\bibfnamefont {S.~L.}\ \bibnamefont {Price}}, \bibinfo {author} {\bibfnamefont {M.~R.}\ \bibnamefont {Probert}}, \bibinfo {author} {\bibfnamefont {A.}~\bibnamefont {Pulido}}, \bibinfo {author} {\bibfnamefont {G.~R.}\ \bibnamefont {Ramteke}}, \bibinfo {author} {\bibfnamefont {A.~U.}\ \bibnamefont {Rehman}}, \bibinfo {author} {\bibfnamefont {S.~M.}\ \bibnamefont {Reutzel-Edens}}, \bibinfo {author} {\bibfnamefont {J.}~\bibnamefont {Rogal}}, \bibinfo {author} {\bibfnamefont {M.~J.}\ \bibnamefont {Ross}}, \bibinfo {author} {\bibfnamefont {A.~F.}\ \bibnamefont {Rumson}}, \bibinfo {author} {\bibfnamefont {G.}~\bibnamefont {Sadiq}}, \bibinfo {author} {\bibfnamefont {Z.~M.}\ \bibnamefont {Saeed}}, \bibinfo {author} {\bibfnamefont {A.}~\bibnamefont {Salimi}}, \bibinfo {author} {\bibfnamefont {K.}~\bibnamefont {Sasikumar}}, \bibinfo {author} {\bibfnamefont {S.}~\bibnamefont
  {Sekharan}}, \bibinfo {author} {\bibfnamefont {K.}~\bibnamefont {Shankland}}, \bibinfo {author} {\bibfnamefont {B.}~\bibnamefont {Shi}}, \bibinfo {author} {\bibfnamefont {X.}~\bibnamefont {Shi}}, \bibinfo {author} {\bibfnamefont {K.}~\bibnamefont {Shinohara}}, \bibinfo {author} {\bibfnamefont {A.~G.}\ \bibnamefont {Skillman}}, \bibinfo {author} {\bibfnamefont {H.}~\bibnamefont {Song}}, \bibinfo {author} {\bibfnamefont {N.}~\bibnamefont {Strasser}}, \bibinfo {author} {\bibfnamefont {J.}~\bibnamefont {van~de Streek}}, \bibinfo {author} {\bibfnamefont {I.~J.}\ \bibnamefont {Sugden}}, \bibinfo {author} {\bibfnamefont {G.}~\bibnamefont {Sun}}, \bibinfo {author} {\bibfnamefont {K.}~\bibnamefont {Szalewicz}}, \bibinfo {author} {\bibfnamefont {L.}~\bibnamefont {Tan}}, \bibinfo {author} {\bibfnamefont {K.}~\bibnamefont {Tang}}, \bibinfo {author} {\bibfnamefont {F.}~\bibnamefont {Tarczynski}}, \bibinfo {author} {\bibfnamefont {C.~R.}\ \bibnamefont {Taylor}}, \bibinfo {author} {\bibfnamefont {A.}~\bibnamefont
  {Tkatchenko}}, \bibinfo {author} {\bibfnamefont {R.}~\bibnamefont {Tom}}, \bibinfo {author} {\bibfnamefont {P.}~\bibnamefont {Touš}}, \bibinfo {author} {\bibfnamefont {M.~E.}\ \bibnamefont {Tuckerman}}, \bibinfo {author} {\bibfnamefont {P.~A.}\ \bibnamefont {Unzueta}}, \bibinfo {author} {\bibfnamefont {Y.}~\bibnamefont {Utsumi}}, \bibinfo {author} {\bibfnamefont {L.}~\bibnamefont {Vogt-Maranto}}, \bibinfo {author} {\bibfnamefont {J.}~\bibnamefont {Weatherston}}, \bibinfo {author} {\bibfnamefont {L.~J.}\ \bibnamefont {Wilkinson}}, \bibinfo {author} {\bibfnamefont {R.~D.}\ \bibnamefont {Willacy}}, \bibinfo {author} {\bibfnamefont {L.}~\bibnamefont {Wojtas}}, \bibinfo {author} {\bibfnamefont {G.~R.}\ \bibnamefont {Woollam}}, \bibinfo {author} {\bibfnamefont {Y.}~\bibnamefont {Yang}}, \bibinfo {author} {\bibfnamefont {Z.}~\bibnamefont {Yang}}, \bibinfo {author} {\bibfnamefont {E.}~\bibnamefont {Yonemochi}}, \bibinfo {author} {\bibfnamefont {X.}~\bibnamefont {Yue}}, \bibinfo {author} {\bibfnamefont
  {Q.}~\bibnamefont {Zeng}}, \bibinfo {author} {\bibfnamefont {T.}~\bibnamefont {Zhou}}, \bibinfo {author} {\bibfnamefont {Y.}~\bibnamefont {Zhou}}, \bibinfo {author} {\bibfnamefont {R.}~\bibnamefont {Zubatyuk}}, \ and\ \bibinfo {author} {\bibfnamefont {J.~C.}\ \bibnamefont {Cole}},\ }\href {\doibase 10.1107/s2052520624008679} {\bibfield  {journal} {\bibinfo  {journal} {Acta Crystallogr. Sect. B: Struct. Sci.}\ }\textbf {\bibinfo {volume} {80}},\ \bibinfo {pages} {548–574} (\bibinfo {year} {2024}{\natexlab{b}})}\BibitemShut {NoStop}%
\bibitem [{\citenamefont {Zhou}\ \emph {et~al.}(2025)\citenamefont {Zhou}, \citenamefont {Bier}, \citenamefont {Santra}, \citenamefont {Jacobson}, \citenamefont {Wu}, \citenamefont {Garaizar~Suarez}, \citenamefont {Almaguer}, \citenamefont {Yu}, \citenamefont {Abel}, \citenamefont {Friesner},\ and\ \citenamefont {Wang}}]{Zhou2025}%
  \BibitemOpen
  \bibfield  {author} {\bibinfo {author} {\bibfnamefont {D.}~\bibnamefont {Zhou}}, \bibinfo {author} {\bibfnamefont {I.}~\bibnamefont {Bier}}, \bibinfo {author} {\bibfnamefont {B.}~\bibnamefont {Santra}}, \bibinfo {author} {\bibfnamefont {L.~D.}\ \bibnamefont {Jacobson}}, \bibinfo {author} {\bibfnamefont {C.}~\bibnamefont {Wu}}, \bibinfo {author} {\bibfnamefont {A.}~\bibnamefont {Garaizar~Suarez}}, \bibinfo {author} {\bibfnamefont {B.~R.}\ \bibnamefont {Almaguer}}, \bibinfo {author} {\bibfnamefont {H.}~\bibnamefont {Yu}}, \bibinfo {author} {\bibfnamefont {R.}~\bibnamefont {Abel}}, \bibinfo {author} {\bibfnamefont {R.~A.}\ \bibnamefont {Friesner}}, \ and\ \bibinfo {author} {\bibfnamefont {L.}~\bibnamefont {Wang}},\ }\href {\doibase 10.1038/s41467-025-57479-1} {\bibfield  {journal} {\bibinfo  {journal} {Nat. Commun.}\ }\textbf {\bibinfo {volume} {16}},\ \bibinfo {pages} {2210} (\bibinfo {year} {2025})}\BibitemShut {NoStop}%
\bibitem [{\citenamefont {Firaha}\ \emph {et~al.}(2023)\citenamefont {Firaha}, \citenamefont {Liu}, \citenamefont {van~de Streek}, \citenamefont {Sasikumar}, \citenamefont {Dietrich}, \citenamefont {Helfferich}, \citenamefont {Aerts}, \citenamefont {Braun}, \citenamefont {Broo}, \citenamefont {DiPasquale}, \citenamefont {Lee}, \citenamefont {Le~Meur}, \citenamefont {Nilsson~Lill}, \citenamefont {Lunsmann}, \citenamefont {Mattei}, \citenamefont {Muglia}, \citenamefont {Putra}, \citenamefont {Raoui}, \citenamefont {Reutzel-Edens}, \citenamefont {Rome}, \citenamefont {Sheikh}, \citenamefont {Tkatchenko}, \citenamefont {Woollam},\ and\ \citenamefont {Neumann}}]{Firaha2023}%
  \BibitemOpen
  \bibfield  {author} {\bibinfo {author} {\bibfnamefont {D.}~\bibnamefont {Firaha}}, \bibinfo {author} {\bibfnamefont {Y.~M.}\ \bibnamefont {Liu}}, \bibinfo {author} {\bibfnamefont {J.}~\bibnamefont {van~de Streek}}, \bibinfo {author} {\bibfnamefont {K.}~\bibnamefont {Sasikumar}}, \bibinfo {author} {\bibfnamefont {H.}~\bibnamefont {Dietrich}}, \bibinfo {author} {\bibfnamefont {J.}~\bibnamefont {Helfferich}}, \bibinfo {author} {\bibfnamefont {L.}~\bibnamefont {Aerts}}, \bibinfo {author} {\bibfnamefont {D.~E.}\ \bibnamefont {Braun}}, \bibinfo {author} {\bibfnamefont {A.}~\bibnamefont {Broo}}, \bibinfo {author} {\bibfnamefont {A.~G.}\ \bibnamefont {DiPasquale}}, \bibinfo {author} {\bibfnamefont {A.~Y.}\ \bibnamefont {Lee}}, \bibinfo {author} {\bibfnamefont {S.}~\bibnamefont {Le~Meur}}, \bibinfo {author} {\bibfnamefont {S.~O.}\ \bibnamefont {Nilsson~Lill}}, \bibinfo {author} {\bibfnamefont {W.~J.}\ \bibnamefont {Lunsmann}}, \bibinfo {author} {\bibfnamefont {A.}~\bibnamefont {Mattei}}, \bibinfo {author}
  {\bibfnamefont {P.}~\bibnamefont {Muglia}}, \bibinfo {author} {\bibfnamefont {O.~D.}\ \bibnamefont {Putra}}, \bibinfo {author} {\bibfnamefont {M.}~\bibnamefont {Raoui}}, \bibinfo {author} {\bibfnamefont {S.~M.}\ \bibnamefont {Reutzel-Edens}}, \bibinfo {author} {\bibfnamefont {S.}~\bibnamefont {Rome}}, \bibinfo {author} {\bibfnamefont {A.~Y.}\ \bibnamefont {Sheikh}}, \bibinfo {author} {\bibfnamefont {A.}~\bibnamefont {Tkatchenko}}, \bibinfo {author} {\bibfnamefont {G.~R.}\ \bibnamefont {Woollam}}, \ and\ \bibinfo {author} {\bibfnamefont {M.~A.}\ \bibnamefont {Neumann}},\ }\href {\doibase 10.1038/s41586-023-06587-3} {\bibfield  {journal} {\bibinfo  {journal} {Nature}\ }\textbf {\bibinfo {volume} {623}},\ \bibinfo {pages} {324–328} (\bibinfo {year} {2023})}\BibitemShut {NoStop}%
\bibitem [{\citenamefont {Butler}\ and\ \citenamefont {Day}(2023)}]{Butler2023}%
  \BibitemOpen
  \bibfield  {author} {\bibinfo {author} {\bibfnamefont {P.~W.~V.}\ \bibnamefont {Butler}}\ and\ \bibinfo {author} {\bibfnamefont {G.~M.}\ \bibnamefont {Day}},\ }\href {\doibase 10.1073/pnas.2300516120} {\bibfield  {journal} {\bibinfo  {journal} {Proc. Natl. Acad. Sci.}\ }\textbf {\bibinfo {volume} {120}},\ \bibinfo {pages} {e2300516120} (\bibinfo {year} {2023})}\BibitemShut {NoStop}%
\bibitem [{\citenamefont {Khorasani}\ \emph {et~al.}(2015)\citenamefont {Khorasani}, \citenamefont {Botes}, \citenamefont {Fernandes},\ and\ \citenamefont {Levendis}}]{Khorasani2015}%
  \BibitemOpen
  \bibfield  {author} {\bibinfo {author} {\bibfnamefont {S.}~\bibnamefont {Khorasani}}, \bibinfo {author} {\bibfnamefont {D.~S.}\ \bibnamefont {Botes}}, \bibinfo {author} {\bibfnamefont {M.~A.}\ \bibnamefont {Fernandes}}, \ and\ \bibinfo {author} {\bibfnamefont {D.~C.}\ \bibnamefont {Levendis}},\ }\href {\doibase 10.1039/c5ce01301a} {\bibfield  {journal} {\bibinfo  {journal} {CrystEngComm}\ }\textbf {\bibinfo {volume} {17}},\ \bibinfo {pages} {8933–8945} (\bibinfo {year} {2015})}\BibitemShut {NoStop}%
\bibitem [{\citenamefont {Nyman}\ and\ \citenamefont {Day}(2015)}]{Nyman2015}%
  \BibitemOpen
  \bibfield  {author} {\bibinfo {author} {\bibfnamefont {J.}~\bibnamefont {Nyman}}\ and\ \bibinfo {author} {\bibfnamefont {G.~M.}\ \bibnamefont {Day}},\ }\href {\doibase 10.1039/c5ce00045a} {\bibfield  {journal} {\bibinfo  {journal} {CrystEngComm}\ }\textbf {\bibinfo {volume} {17}},\ \bibinfo {pages} {5154–5165} (\bibinfo {year} {2015})}\BibitemShut {NoStop}%
\bibitem [{\citenamefont {Hoja}\ \emph {et~al.}(2019)\citenamefont {Hoja}, \citenamefont {Ko}, \citenamefont {Neumann}, \citenamefont {Car}, \citenamefont {DiStasio},\ and\ \citenamefont {Tkatchenko}}]{Hoja2019}%
  \BibitemOpen
  \bibfield  {author} {\bibinfo {author} {\bibfnamefont {J.}~\bibnamefont {Hoja}}, \bibinfo {author} {\bibfnamefont {H.-Y.}\ \bibnamefont {Ko}}, \bibinfo {author} {\bibfnamefont {M.~A.}\ \bibnamefont {Neumann}}, \bibinfo {author} {\bibfnamefont {R.}~\bibnamefont {Car}}, \bibinfo {author} {\bibfnamefont {R.~A.}\ \bibnamefont {DiStasio}}, \ and\ \bibinfo {author} {\bibfnamefont {A.}~\bibnamefont {Tkatchenko}},\ }\href {\doibase 10.1126/sciadv.aau3338} {\bibfield  {journal} {\bibinfo  {journal} {Sci. Adv.}\ }\textbf {\bibinfo {volume} {5}},\ \bibinfo {pages} {eaau3338} (\bibinfo {year} {2019})}\BibitemShut {NoStop}%
\bibitem [{\citenamefont {Weatherby}\ \emph {et~al.}(2022)\citenamefont {Weatherby}, \citenamefont {Rumson}, \citenamefont {Price}, \citenamefont {Otero de~la Roza},\ and\ \citenamefont {Johnson}}]{Weatherby2022}%
  \BibitemOpen
  \bibfield  {author} {\bibinfo {author} {\bibfnamefont {J.~A.}\ \bibnamefont {Weatherby}}, \bibinfo {author} {\bibfnamefont {A.~F.}\ \bibnamefont {Rumson}}, \bibinfo {author} {\bibfnamefont {A.~J.~A.}\ \bibnamefont {Price}}, \bibinfo {author} {\bibfnamefont {A.}~\bibnamefont {Otero de~la Roza}}, \ and\ \bibinfo {author} {\bibfnamefont {E.~R.}\ \bibnamefont {Johnson}},\ }\href {\doibase 10.1063/5.0083082} {\bibfield  {journal} {\bibinfo  {journal} {J. Chem. Phys.}\ }\textbf {\bibinfo {volume} {156}},\ \bibinfo {pages} {114108} (\bibinfo {year} {2022})}\BibitemShut {NoStop}%
\bibitem [{\citenamefont {Kapil}\ and\ \citenamefont {Engel}(2022)}]{Kapil2022}%
  \BibitemOpen
  \bibfield  {author} {\bibinfo {author} {\bibfnamefont {V.}~\bibnamefont {Kapil}}\ and\ \bibinfo {author} {\bibfnamefont {E.~A.}\ \bibnamefont {Engel}},\ }\href {\doibase 10.1073/pnas.2111769119} {\bibfield  {journal} {\bibinfo  {journal} {Proc. Natl. Acad. Sci. U.S.A.}\ }\textbf {\bibinfo {volume} {119}},\ \bibinfo {pages} {e2111769119} (\bibinfo {year} {2022})}\BibitemShut {NoStop}%
\bibitem [{\citenamefont {Nyman}\ and\ \citenamefont {Day}(2016)}]{Nyman2016}%
  \BibitemOpen
  \bibfield  {author} {\bibinfo {author} {\bibfnamefont {J.}~\bibnamefont {Nyman}}\ and\ \bibinfo {author} {\bibfnamefont {G.~M.}\ \bibnamefont {Day}},\ }\href {\doibase 10.1039/c6cp05447a} {\bibfield  {journal} {\bibinfo  {journal} {Phys. Chem. Chem. Phys.}\ }\textbf {\bibinfo {volume} {18}},\ \bibinfo {pages} {31132–31143} (\bibinfo {year} {2016})}\BibitemShut {NoStop}%
\bibitem [{\citenamefont {Raimbault}, \citenamefont {Athavale},\ and\ \citenamefont {Rossi}(2019)}]{Raimbault2019b}%
  \BibitemOpen
  \bibfield  {author} {\bibinfo {author} {\bibfnamefont {N.}~\bibnamefont {Raimbault}}, \bibinfo {author} {\bibfnamefont {V.}~\bibnamefont {Athavale}}, \ and\ \bibinfo {author} {\bibfnamefont {M.}~\bibnamefont {Rossi}},\ }\href {\doibase 10.1103/physrevmaterials.3.053605} {\bibfield  {journal} {\bibinfo  {journal} {Phys. Rev. Mater.}\ }\textbf {\bibinfo {volume} {3}},\ \bibinfo {pages} {053605} (\bibinfo {year} {2019})}\BibitemShut {NoStop}%
\bibitem [{\citenamefont {Dolgonos}, \citenamefont {Hoja},\ and\ \citenamefont {Boese}(2019)}]{Dolgonos2019}%
  \BibitemOpen
  \bibfield  {author} {\bibinfo {author} {\bibfnamefont {G.~A.}\ \bibnamefont {Dolgonos}}, \bibinfo {author} {\bibfnamefont {J.}~\bibnamefont {Hoja}}, \ and\ \bibinfo {author} {\bibfnamefont {A.~D.}\ \bibnamefont {Boese}},\ }\href {\doibase 10.1039/c9cp04488d} {\bibfield  {journal} {\bibinfo  {journal} {Phys. Chem. Chem. Phys.}\ }\textbf {\bibinfo {volume} {21}},\ \bibinfo {pages} {24333} (\bibinfo {year} {2019})}\BibitemShut {NoStop}%
\bibitem [{\citenamefont {Hattori}\ and\ \citenamefont {Zhu}(2024)}]{Hattori2024}%
  \BibitemOpen
  \bibfield  {author} {\bibinfo {author} {\bibfnamefont {S.}~\bibnamefont {Hattori}}\ and\ \bibinfo {author} {\bibfnamefont {Q.}~\bibnamefont {Zhu}},\ }\href {\doibase 10.1021/acsomega.4c04782} {\bibfield  {journal} {\bibinfo  {journal} {ACS Omega}\ }\textbf {\bibinfo {volume} {9}},\ \bibinfo {pages} {36589–36599} (\bibinfo {year} {2024})}\BibitemShut {NoStop}%
\bibitem [{\citenamefont {Musil}\ \emph {et~al.}(2018)\citenamefont {Musil}, \citenamefont {De}, \citenamefont {Yang}, \citenamefont {Campbell}, \citenamefont {Day},\ and\ \citenamefont {Ceriotti}}]{Musil2018}%
  \BibitemOpen
  \bibfield  {author} {\bibinfo {author} {\bibfnamefont {F.}~\bibnamefont {Musil}}, \bibinfo {author} {\bibfnamefont {S.}~\bibnamefont {De}}, \bibinfo {author} {\bibfnamefont {J.}~\bibnamefont {Yang}}, \bibinfo {author} {\bibfnamefont {J.~E.}\ \bibnamefont {Campbell}}, \bibinfo {author} {\bibfnamefont {G.~M.}\ \bibnamefont {Day}}, \ and\ \bibinfo {author} {\bibfnamefont {M.}~\bibnamefont {Ceriotti}},\ }\href {\doibase 10.1039/c7sc04665k} {\bibfield  {journal} {\bibinfo  {journal} {Chem. Sci.}\ }\textbf {\bibinfo {volume} {9}},\ \bibinfo {pages} {1289–1300} (\bibinfo {year} {2018})}\BibitemShut {NoStop}%
\bibitem [{\citenamefont {Raimbault}\ \emph {et~al.}(2019)\citenamefont {Raimbault}, \citenamefont {Grisafi}, \citenamefont {Ceriotti},\ and\ \citenamefont {Rossi}}]{Raimbault2019}%
  \BibitemOpen
  \bibfield  {author} {\bibinfo {author} {\bibfnamefont {N.}~\bibnamefont {Raimbault}}, \bibinfo {author} {\bibfnamefont {A.}~\bibnamefont {Grisafi}}, \bibinfo {author} {\bibfnamefont {M.}~\bibnamefont {Ceriotti}}, \ and\ \bibinfo {author} {\bibfnamefont {M.}~\bibnamefont {Rossi}},\ }\href {\doibase 10.1088/1367-2630/ab4509} {\bibfield  {journal} {\bibinfo  {journal} {New J. Phys.}\ }\textbf {\bibinfo {volume} {21}},\ \bibinfo {pages} {105001} (\bibinfo {year} {2019})}\BibitemShut {NoStop}%
\bibitem [{\citenamefont {Krynski}\ and\ \citenamefont {Rossi}(2021)}]{Krynski2021}%
  \BibitemOpen
  \bibfield  {author} {\bibinfo {author} {\bibfnamefont {M.}~\bibnamefont {Krynski}}\ and\ \bibinfo {author} {\bibfnamefont {M.}~\bibnamefont {Rossi}},\ }\href {\doibase 10.1038/s41524-021-00638-x} {\bibfield  {journal} {\bibinfo  {journal} {Npj Comput. Mater.}\ }\textbf {\bibinfo {volume} {7}},\ \bibinfo {pages} {169} (\bibinfo {year} {2021})}\BibitemShut {NoStop}%
\bibitem [{\citenamefont {Della~Pia}\ \emph {et~al.}(2025)\citenamefont {Della~Pia}, \citenamefont {Shi}, \citenamefont {Kapil}, \citenamefont {Zen}, \citenamefont {Alfè},\ and\ \citenamefont {Michaelides}}]{DellaPia2025}%
  \BibitemOpen
  \bibfield  {author} {\bibinfo {author} {\bibfnamefont {F.}~\bibnamefont {Della~Pia}}, \bibinfo {author} {\bibfnamefont {B.~X.}\ \bibnamefont {Shi}}, \bibinfo {author} {\bibfnamefont {V.}~\bibnamefont {Kapil}}, \bibinfo {author} {\bibfnamefont {A.}~\bibnamefont {Zen}}, \bibinfo {author} {\bibfnamefont {D.}~\bibnamefont {Alfè}}, \ and\ \bibinfo {author} {\bibfnamefont {A.}~\bibnamefont {Michaelides}},\ }\href {\doibase 10.1039/d5sc01325a} {\bibfield  {journal} {\bibinfo  {journal} {Chem. Sci.}\ }\textbf {\bibinfo {volume} {16}},\ \bibinfo {pages} {11419–11433} (\bibinfo {year} {2025})}\BibitemShut {NoStop}%
\bibitem [{\citenamefont {Gurlek}\ \emph {et~al.}(2025)\citenamefont {Gurlek}, \citenamefont {Sharma}, \citenamefont {Lazzaroni}, \citenamefont {Rubio},\ and\ \citenamefont {Rossi}}]{Gurlek2025}%
  \BibitemOpen
  \bibfield  {author} {\bibinfo {author} {\bibfnamefont {B.}~\bibnamefont {Gurlek}}, \bibinfo {author} {\bibfnamefont {S.}~\bibnamefont {Sharma}}, \bibinfo {author} {\bibfnamefont {P.}~\bibnamefont {Lazzaroni}}, \bibinfo {author} {\bibfnamefont {A.}~\bibnamefont {Rubio}}, \ and\ \bibinfo {author} {\bibfnamefont {M.}~\bibnamefont {Rossi}},\ }\href {\doibase 10.1038/s41524-025-01825-w} {\bibfield  {journal} {\bibinfo  {journal} {npj Comput. Mater.}\ }\textbf {\bibinfo {volume} {11}},\ \bibinfo {pages} {318} (\bibinfo {year} {2025})}\BibitemShut {NoStop}%
\bibitem [{\citenamefont {Kabylda}\ \emph {et~al.}(2025)\citenamefont {Kabylda}, \citenamefont {Frank}, \citenamefont {Suárez-Dou}, \citenamefont {Khabibrakhmanov}, \citenamefont {Medrano~Sandonas}, \citenamefont {Unke}, \citenamefont {Chmiela}, \citenamefont {M\"{u}ller},\ and\ \citenamefont {Tkatchenko}}]{Kabylda2025}%
  \BibitemOpen
  \bibfield  {author} {\bibinfo {author} {\bibfnamefont {A.}~\bibnamefont {Kabylda}}, \bibinfo {author} {\bibfnamefont {J.~T.}\ \bibnamefont {Frank}}, \bibinfo {author} {\bibfnamefont {S.}~\bibnamefont {Suárez-Dou}}, \bibinfo {author} {\bibfnamefont {A.}~\bibnamefont {Khabibrakhmanov}}, \bibinfo {author} {\bibfnamefont {L.}~\bibnamefont {Medrano~Sandonas}}, \bibinfo {author} {\bibfnamefont {O.~T.}\ \bibnamefont {Unke}}, \bibinfo {author} {\bibfnamefont {S.}~\bibnamefont {Chmiela}}, \bibinfo {author} {\bibfnamefont {K.-R.}\ \bibnamefont {M\"{u}ller}}, \ and\ \bibinfo {author} {\bibfnamefont {A.}~\bibnamefont {Tkatchenko}},\ }\href {\doibase 10.1021/jacs.5c09558} {\bibfield  {journal} {\bibinfo  {journal} {J. Am. Chem. Soc.}\ }\textbf {\bibinfo {volume} {147}},\ \bibinfo {pages} {33723–33734} (\bibinfo {year} {2025})}\BibitemShut {NoStop}%
\bibitem [{\citenamefont {Kaur}\ \emph {et~al.}(2025)\citenamefont {Kaur}, \citenamefont {Della~Pia}, \citenamefont {Batatia}, \citenamefont {Advincula}, \citenamefont {Shi}, \citenamefont {Lan}, \citenamefont {Csányi}, \citenamefont {Michaelides},\ and\ \citenamefont {Kapil}}]{Kaur2025}%
  \BibitemOpen
  \bibfield  {author} {\bibinfo {author} {\bibfnamefont {H.}~\bibnamefont {Kaur}}, \bibinfo {author} {\bibfnamefont {F.}~\bibnamefont {Della~Pia}}, \bibinfo {author} {\bibfnamefont {I.}~\bibnamefont {Batatia}}, \bibinfo {author} {\bibfnamefont {X.~R.}\ \bibnamefont {Advincula}}, \bibinfo {author} {\bibfnamefont {B.~X.}\ \bibnamefont {Shi}}, \bibinfo {author} {\bibfnamefont {J.}~\bibnamefont {Lan}}, \bibinfo {author} {\bibfnamefont {G.}~\bibnamefont {Csányi}}, \bibinfo {author} {\bibfnamefont {A.}~\bibnamefont {Michaelides}}, \ and\ \bibinfo {author} {\bibfnamefont {V.}~\bibnamefont {Kapil}},\ }\href {\doibase 10.1039/d4fd00107a} {\bibfield  {journal} {\bibinfo  {journal} {Faraday Discuss.}\ }\textbf {\bibinfo {volume} {256}},\ \bibinfo {pages} {120–138} (\bibinfo {year} {2025})}\BibitemShut {NoStop}%
\bibitem [{\citenamefont {Goncharova}\ and\ \citenamefont {Hoja}(2025)}]{Goncharova2025}%
  \BibitemOpen
  \bibfield  {author} {\bibinfo {author} {\bibfnamefont {N.}~\bibnamefont {Goncharova}}\ and\ \bibinfo {author} {\bibfnamefont {J.}~\bibnamefont {Hoja}},\ }\href {\doibase 10.1063/5.0285417} {\bibfield  {journal} {\bibinfo  {journal} {J. Chem. Phys.}\ }\textbf {\bibinfo {volume} {163}},\ \bibinfo {pages} {164127} (\bibinfo {year} {2025})}\BibitemShut {NoStop}%
\bibitem [{\citenamefont {Alcob{\'{e}}}\ \emph {et~al.}(1994)\citenamefont {Alcob{\'{e}}}, \citenamefont {Estop}, \citenamefont {Aliev}, \citenamefont {Harris}, \citenamefont {Rodriguez-Carvajal},\ and\ \citenamefont {Rius}}]{Alcob1994}%
  \BibitemOpen
  \bibfield  {author} {\bibinfo {author} {\bibfnamefont {X.}~\bibnamefont {Alcob{\'{e}}}}, \bibinfo {author} {\bibfnamefont {E.}~\bibnamefont {Estop}}, \bibinfo {author} {\bibfnamefont {A.~E.}\ \bibnamefont {Aliev}}, \bibinfo {author} {\bibfnamefont {K.~D.}\ \bibnamefont {Harris}}, \bibinfo {author} {\bibfnamefont {J.}~\bibnamefont {Rodriguez-Carvajal}}, \ and\ \bibinfo {author} {\bibfnamefont {J.}~\bibnamefont {Rius}},\ }\href {\doibase 10.1006/jssc.1994.1129} {\bibfield  {journal} {\bibinfo  {journal} {J. Solid State Chem.}\ }\textbf {\bibinfo {volume} {110}},\ \bibinfo {pages} {20} (\bibinfo {year} {1994})}\BibitemShut {NoStop}%
\bibitem [{\citenamefont {Blum}\ \emph {et~al.}(2009)\citenamefont {Blum}, \citenamefont {Gehrke}, \citenamefont {Hanke}, \citenamefont {Havu}, \citenamefont {Havu}, \citenamefont {Ren}, \citenamefont {Reuter},\ and\ \citenamefont {Scheffler}}]{Blum2009}%
  \BibitemOpen
  \bibfield  {author} {\bibinfo {author} {\bibfnamefont {V.}~\bibnamefont {Blum}}, \bibinfo {author} {\bibfnamefont {R.}~\bibnamefont {Gehrke}}, \bibinfo {author} {\bibfnamefont {F.}~\bibnamefont {Hanke}}, \bibinfo {author} {\bibfnamefont {P.}~\bibnamefont {Havu}}, \bibinfo {author} {\bibfnamefont {V.}~\bibnamefont {Havu}}, \bibinfo {author} {\bibfnamefont {X.}~\bibnamefont {Ren}}, \bibinfo {author} {\bibfnamefont {K.}~\bibnamefont {Reuter}}, \ and\ \bibinfo {author} {\bibfnamefont {M.}~\bibnamefont {Scheffler}},\ }\href {\doibase 10.1016/j.cpc.2009.06.022} {\bibfield  {journal} {\bibinfo  {journal} {Comput. Phys. Commun.}\ }\textbf {\bibinfo {volume} {180}},\ \bibinfo {pages} {2175} (\bibinfo {year} {2009})}\BibitemShut {NoStop}%
\bibitem [{\citenamefont {Knuth}\ \emph {et~al.}(2015)\citenamefont {Knuth}, \citenamefont {Carbogno}, \citenamefont {Atalla}, \citenamefont {Blum},\ and\ \citenamefont {Scheffler}}]{Knuth2015}%
  \BibitemOpen
  \bibfield  {author} {\bibinfo {author} {\bibfnamefont {F.}~\bibnamefont {Knuth}}, \bibinfo {author} {\bibfnamefont {C.}~\bibnamefont {Carbogno}}, \bibinfo {author} {\bibfnamefont {V.}~\bibnamefont {Atalla}}, \bibinfo {author} {\bibfnamefont {V.}~\bibnamefont {Blum}}, \ and\ \bibinfo {author} {\bibfnamefont {M.}~\bibnamefont {Scheffler}},\ }\href {\doibase 10.1016/j.cpc.2015.01.003} {\bibfield  {journal} {\bibinfo  {journal} {Comput. Phys. Commun.}\ }\textbf {\bibinfo {volume} {190}},\ \bibinfo {pages} {33} (\bibinfo {year} {2015})}\BibitemShut {NoStop}%
\bibitem [{\citenamefont {Ren}\ \emph {et~al.}(2012)\citenamefont {Ren}, \citenamefont {Rinke}, \citenamefont {Blum}, \citenamefont {Wieferink}, \citenamefont {Tkatchenko}, \citenamefont {Sanfilippo}, \citenamefont {Reuter},\ and\ \citenamefont {Scheffler}}]{Ren2012}%
  \BibitemOpen
  \bibfield  {author} {\bibinfo {author} {\bibfnamefont {X.}~\bibnamefont {Ren}}, \bibinfo {author} {\bibfnamefont {P.}~\bibnamefont {Rinke}}, \bibinfo {author} {\bibfnamefont {V.}~\bibnamefont {Blum}}, \bibinfo {author} {\bibfnamefont {J.}~\bibnamefont {Wieferink}}, \bibinfo {author} {\bibfnamefont {A.}~\bibnamefont {Tkatchenko}}, \bibinfo {author} {\bibfnamefont {A.}~\bibnamefont {Sanfilippo}}, \bibinfo {author} {\bibfnamefont {K.}~\bibnamefont {Reuter}}, \ and\ \bibinfo {author} {\bibfnamefont {M.}~\bibnamefont {Scheffler}},\ }\href {\doibase 10.1088/1367-2630/14/5/053020} {\bibfield  {journal} {\bibinfo  {journal} {New J. Phys.}\ }\textbf {\bibinfo {volume} {14}},\ \bibinfo {pages} {053020} (\bibinfo {year} {2012})}\BibitemShut {NoStop}%
\bibitem [{\citenamefont {Yu}\ \emph {et~al.}(2018)\citenamefont {Yu}, \citenamefont {Corsetti}, \citenamefont {Garc{\'{\i}}a}, \citenamefont {Huhn}, \citenamefont {Jacquelin}, \citenamefont {Jia}, \citenamefont {Lange}, \citenamefont {Lin}, \citenamefont {Lu}, \citenamefont {Mi}, \citenamefont {Seifitokaldani}, \citenamefont {V{\'{a}}zquez-Mayagoitia}, \citenamefont {Yang}, \citenamefont {Yang},\ and\ \citenamefont {Blum}}]{Yu2018}%
  \BibitemOpen
  \bibfield  {author} {\bibinfo {author} {\bibfnamefont {V.~W.}\ \bibnamefont {Yu}}, \bibinfo {author} {\bibfnamefont {F.}~\bibnamefont {Corsetti}}, \bibinfo {author} {\bibfnamefont {A.}~\bibnamefont {Garc{\'{\i}}a}}, \bibinfo {author} {\bibfnamefont {W.~P.}\ \bibnamefont {Huhn}}, \bibinfo {author} {\bibfnamefont {M.}~\bibnamefont {Jacquelin}}, \bibinfo {author} {\bibfnamefont {W.}~\bibnamefont {Jia}}, \bibinfo {author} {\bibfnamefont {B.}~\bibnamefont {Lange}}, \bibinfo {author} {\bibfnamefont {L.}~\bibnamefont {Lin}}, \bibinfo {author} {\bibfnamefont {J.}~\bibnamefont {Lu}}, \bibinfo {author} {\bibfnamefont {W.}~\bibnamefont {Mi}}, \bibinfo {author} {\bibfnamefont {A.}~\bibnamefont {Seifitokaldani}}, \bibinfo {author} {\bibfnamefont {{\'{A}}.}~\bibnamefont {V{\'{a}}zquez-Mayagoitia}}, \bibinfo {author} {\bibfnamefont {C.}~\bibnamefont {Yang}}, \bibinfo {author} {\bibfnamefont {H.}~\bibnamefont {Yang}}, \ and\ \bibinfo {author} {\bibfnamefont {V.}~\bibnamefont {Blum}},\ }\href {\doibase
  10.1016/j.cpc.2017.09.007} {\bibfield  {journal} {\bibinfo  {journal} {Comput. Phys. Commun.}\ }\textbf {\bibinfo {volume} {222}},\ \bibinfo {pages} {267} (\bibinfo {year} {2018})}\BibitemShut {NoStop}%
\bibitem [{\citenamefont {Havu}\ \emph {et~al.}(2009)\citenamefont {Havu}, \citenamefont {Blum}, \citenamefont {Havu},\ and\ \citenamefont {Scheffler}}]{Havu2009}%
  \BibitemOpen
  \bibfield  {author} {\bibinfo {author} {\bibfnamefont {V.}~\bibnamefont {Havu}}, \bibinfo {author} {\bibfnamefont {V.}~\bibnamefont {Blum}}, \bibinfo {author} {\bibfnamefont {P.}~\bibnamefont {Havu}}, \ and\ \bibinfo {author} {\bibfnamefont {M.}~\bibnamefont {Scheffler}},\ }\href {\doibase 10.1016/j.jcp.2009.08.008} {\bibfield  {journal} {\bibinfo  {journal} {J. Comput. Phys.}\ }\textbf {\bibinfo {volume} {228}},\ \bibinfo {pages} {8367} (\bibinfo {year} {2009})}\BibitemShut {NoStop}%
\bibitem [{\citenamefont {Ihrig}\ \emph {et~al.}(2015)\citenamefont {Ihrig}, \citenamefont {Wieferink}, \citenamefont {Zhang}, \citenamefont {Ropo}, \citenamefont {Ren}, \citenamefont {Rinke}, \citenamefont {Scheffler},\ and\ \citenamefont {Blum}}]{Ihrig2015}%
  \BibitemOpen
  \bibfield  {author} {\bibinfo {author} {\bibfnamefont {A.~C.}\ \bibnamefont {Ihrig}}, \bibinfo {author} {\bibfnamefont {J.}~\bibnamefont {Wieferink}}, \bibinfo {author} {\bibfnamefont {I.~Y.}\ \bibnamefont {Zhang}}, \bibinfo {author} {\bibfnamefont {M.}~\bibnamefont {Ropo}}, \bibinfo {author} {\bibfnamefont {X.}~\bibnamefont {Ren}}, \bibinfo {author} {\bibfnamefont {P.}~\bibnamefont {Rinke}}, \bibinfo {author} {\bibfnamefont {M.}~\bibnamefont {Scheffler}}, \ and\ \bibinfo {author} {\bibfnamefont {V.}~\bibnamefont {Blum}},\ }\href {\doibase 10.1088/1367-2630/17/9/093020} {\bibfield  {journal} {\bibinfo  {journal} {New J. Phys.}\ }\textbf {\bibinfo {volume} {17}},\ \bibinfo {pages} {093020} (\bibinfo {year} {2015})}\BibitemShut {NoStop}%
\bibitem [{\citenamefont {Perdew}, \citenamefont {Burke},\ and\ \citenamefont {Ernzerhof}(1996)}]{Perdew1996}%
  \BibitemOpen
  \bibfield  {author} {\bibinfo {author} {\bibfnamefont {J.~P.}\ \bibnamefont {Perdew}}, \bibinfo {author} {\bibfnamefont {K.}~\bibnamefont {Burke}}, \ and\ \bibinfo {author} {\bibfnamefont {M.}~\bibnamefont {Ernzerhof}},\ }\href {\doibase 10.1103/physrevlett.77.3865} {\bibfield  {journal} {\bibinfo  {journal} {Phys. Rev. Lett.}\ }\textbf {\bibinfo {volume} {77}},\ \bibinfo {pages} {3865} (\bibinfo {year} {1996})}\BibitemShut {NoStop}%
\bibitem [{\citenamefont {Tkatchenko}\ \emph {et~al.}(2012)\citenamefont {Tkatchenko}, \citenamefont {DiStasio}, \citenamefont {Car},\ and\ \citenamefont {Scheffler}}]{Tkatchenko2012}%
  \BibitemOpen
  \bibfield  {author} {\bibinfo {author} {\bibfnamefont {A.}~\bibnamefont {Tkatchenko}}, \bibinfo {author} {\bibfnamefont {R.~A.}\ \bibnamefont {DiStasio}}, \bibinfo {author} {\bibfnamefont {R.}~\bibnamefont {Car}}, \ and\ \bibinfo {author} {\bibfnamefont {M.}~\bibnamefont {Scheffler}},\ }\href {\doibase 10.1103/physrevlett.108.236402} {\bibfield  {journal} {\bibinfo  {journal} {Phys. Rev. Lett.}\ }\textbf {\bibinfo {volume} {108}},\ \bibinfo {pages} {236402} (\bibinfo {year} {2012})}\BibitemShut {NoStop}%
\bibitem [{\citenamefont {Ambrosetti}\ \emph {et~al.}(2014)\citenamefont {Ambrosetti}, \citenamefont {Reilly}, \citenamefont {DiStasio},\ and\ \citenamefont {Tkatchenko}}]{Ambrosetti2014}%
  \BibitemOpen
  \bibfield  {author} {\bibinfo {author} {\bibfnamefont {A.}~\bibnamefont {Ambrosetti}}, \bibinfo {author} {\bibfnamefont {A.~M.}\ \bibnamefont {Reilly}}, \bibinfo {author} {\bibfnamefont {R.~A.}\ \bibnamefont {DiStasio}}, \ and\ \bibinfo {author} {\bibfnamefont {A.}~\bibnamefont {Tkatchenko}},\ }\href {\doibase 10.1063/1.4865104} {\bibfield  {journal} {\bibinfo  {journal} {J. Chem. Phys.}\ }\textbf {\bibinfo {volume} {140}},\ \bibinfo {pages} {18A508} (\bibinfo {year} {2014})}\BibitemShut {NoStop}%
\bibitem [{\citenamefont {Hermann}\ and\ \citenamefont {Tkatchenko}(2020)}]{Hermann2020}%
  \BibitemOpen
  \bibfield  {author} {\bibinfo {author} {\bibfnamefont {J.}~\bibnamefont {Hermann}}\ and\ \bibinfo {author} {\bibfnamefont {A.}~\bibnamefont {Tkatchenko}},\ }\href {\doibase 10.1103/physrevlett.124.146401} {\bibfield  {journal} {\bibinfo  {journal} {Phys. Rev. Lett.}\ }\textbf {\bibinfo {volume} {124}},\ \bibinfo {pages} {146401} (\bibinfo {year} {2020})}\BibitemShut {NoStop}%
\bibitem [{\citenamefont {Adamo}\ and\ \citenamefont {Barone}(1999)}]{Adamo1999}%
  \BibitemOpen
  \bibfield  {author} {\bibinfo {author} {\bibfnamefont {C.}~\bibnamefont {Adamo}}\ and\ \bibinfo {author} {\bibfnamefont {V.}~\bibnamefont {Barone}},\ }\href {\doibase 10.1063/1.478522} {\bibfield  {journal} {\bibinfo  {journal} {The Journal of Chemical Physics}\ }\textbf {\bibinfo {volume} {110}},\ \bibinfo {pages} {6158–6170} (\bibinfo {year} {1999})}\BibitemShut {NoStop}%
\bibitem [{\citenamefont {Larsen}\ \emph {et~al.}(2017)\citenamefont {Larsen}, \citenamefont {Mortensen}, \citenamefont {Blomqvist}, \citenamefont {Castelli}, \citenamefont {Christensen}, \citenamefont {Dułak}, \citenamefont {Friis}, \citenamefont {Groves}, \citenamefont {Hammer}, \citenamefont {Hargus}, \citenamefont {Hermes}, \citenamefont {Jennings}, \citenamefont {Jensen}, \citenamefont {Kermode}, \citenamefont {Kitchin}, \citenamefont {Kolsbjerg}, \citenamefont {Kubal}, \citenamefont {Kaasbjerg}, \citenamefont {Lysgaard}, \citenamefont {Maronsson}, \citenamefont {Maxson}, \citenamefont {Olsen}, \citenamefont {Pastewka}, \citenamefont {Peterson}, \citenamefont {Rostgaard}, \citenamefont {Schiøtz}, \citenamefont {Schütt}, \citenamefont {Strange}, \citenamefont {Thygesen}, \citenamefont {Vegge}, \citenamefont {Vilhelmsen}, \citenamefont {Walter}, \citenamefont {Zeng},\ and\ \citenamefont {Jacobsen}}]{ase-paper}%
  \BibitemOpen
  \bibfield  {author} {\bibinfo {author} {\bibfnamefont {A.~H.}\ \bibnamefont {Larsen}}, \bibinfo {author} {\bibfnamefont {J.~J.}\ \bibnamefont {Mortensen}}, \bibinfo {author} {\bibfnamefont {J.}~\bibnamefont {Blomqvist}}, \bibinfo {author} {\bibfnamefont {I.~E.}\ \bibnamefont {Castelli}}, \bibinfo {author} {\bibfnamefont {R.}~\bibnamefont {Christensen}}, \bibinfo {author} {\bibfnamefont {M.}~\bibnamefont {Dułak}}, \bibinfo {author} {\bibfnamefont {J.}~\bibnamefont {Friis}}, \bibinfo {author} {\bibfnamefont {M.~N.}\ \bibnamefont {Groves}}, \bibinfo {author} {\bibfnamefont {B.}~\bibnamefont {Hammer}}, \bibinfo {author} {\bibfnamefont {C.}~\bibnamefont {Hargus}}, \bibinfo {author} {\bibfnamefont {E.~D.}\ \bibnamefont {Hermes}}, \bibinfo {author} {\bibfnamefont {P.~C.}\ \bibnamefont {Jennings}}, \bibinfo {author} {\bibfnamefont {P.~B.}\ \bibnamefont {Jensen}}, \bibinfo {author} {\bibfnamefont {J.}~\bibnamefont {Kermode}}, \bibinfo {author} {\bibfnamefont {J.~R.}\ \bibnamefont {Kitchin}}, \bibinfo {author}
  {\bibfnamefont {E.~L.}\ \bibnamefont {Kolsbjerg}}, \bibinfo {author} {\bibfnamefont {J.}~\bibnamefont {Kubal}}, \bibinfo {author} {\bibfnamefont {K.}~\bibnamefont {Kaasbjerg}}, \bibinfo {author} {\bibfnamefont {S.}~\bibnamefont {Lysgaard}}, \bibinfo {author} {\bibfnamefont {J.~B.}\ \bibnamefont {Maronsson}}, \bibinfo {author} {\bibfnamefont {T.}~\bibnamefont {Maxson}}, \bibinfo {author} {\bibfnamefont {T.}~\bibnamefont {Olsen}}, \bibinfo {author} {\bibfnamefont {L.}~\bibnamefont {Pastewka}}, \bibinfo {author} {\bibfnamefont {A.}~\bibnamefont {Peterson}}, \bibinfo {author} {\bibfnamefont {C.}~\bibnamefont {Rostgaard}}, \bibinfo {author} {\bibfnamefont {J.}~\bibnamefont {Schiøtz}}, \bibinfo {author} {\bibfnamefont {O.}~\bibnamefont {Schütt}}, \bibinfo {author} {\bibfnamefont {M.}~\bibnamefont {Strange}}, \bibinfo {author} {\bibfnamefont {K.~S.}\ \bibnamefont {Thygesen}}, \bibinfo {author} {\bibfnamefont {T.}~\bibnamefont {Vegge}}, \bibinfo {author} {\bibfnamefont {L.}~\bibnamefont {Vilhelmsen}}, \bibinfo
  {author} {\bibfnamefont {M.}~\bibnamefont {Walter}}, \bibinfo {author} {\bibfnamefont {Z.}~\bibnamefont {Zeng}}, \ and\ \bibinfo {author} {\bibfnamefont {K.~W.}\ \bibnamefont {Jacobsen}},\ }\href {http://stacks.iop.org/0953-8984/29/i=27/a=273002} {\bibfield  {journal} {\bibinfo  {journal} {J. Phys.: Condens. Matter}\ }\textbf {\bibinfo {volume} {29}},\ \bibinfo {pages} {273002} (\bibinfo {year} {2017})}\BibitemShut {NoStop}%
\bibitem [{\citenamefont {Togo}\ and\ \citenamefont {Tanaka}(2015)}]{Togo2015}%
  \BibitemOpen
  \bibfield  {author} {\bibinfo {author} {\bibfnamefont {A.}~\bibnamefont {Togo}}\ and\ \bibinfo {author} {\bibfnamefont {I.}~\bibnamefont {Tanaka}},\ }\href {\doibase 10.1016/j.scriptamat.2015.07.021} {\bibfield  {journal} {\bibinfo  {journal} {Scr. Mater.}\ }\textbf {\bibinfo {volume} {108}},\ \bibinfo {pages} {1} (\bibinfo {year} {2015})}\BibitemShut {NoStop}%
\bibitem [{\citenamefont {Togo}\ \emph {et~al.}(2023)\citenamefont {Togo}, \citenamefont {Chaput}, \citenamefont {Tadano},\ and\ \citenamefont {Tanaka}}]{phonopy-phono3py-JPCM}%
  \BibitemOpen
  \bibfield  {author} {\bibinfo {author} {\bibfnamefont {A.}~\bibnamefont {Togo}}, \bibinfo {author} {\bibfnamefont {L.}~\bibnamefont {Chaput}}, \bibinfo {author} {\bibfnamefont {T.}~\bibnamefont {Tadano}}, \ and\ \bibinfo {author} {\bibfnamefont {I.}~\bibnamefont {Tanaka}},\ }\href {\doibase 10.1088/1361-648X/acd831} {\bibfield  {journal} {\bibinfo  {journal} {J. Phys.: Condens.Matter}\ }\textbf {\bibinfo {volume} {35}},\ \bibinfo {pages} {353001} (\bibinfo {year} {2023})}\BibitemShut {NoStop}%
\bibitem [{\citenamefont {Crouse}(2016)}]{Crouse2016}%
  \BibitemOpen
  \bibfield  {author} {\bibinfo {author} {\bibfnamefont {D.~F.}\ \bibnamefont {Crouse}},\ }\href {\doibase 10.1109/taes.2016.140952} {\bibfield  {journal} {\bibinfo  {journal} {IEEE Transactions on Aerospace and Electronic Systems}\ }\textbf {\bibinfo {volume} {52}},\ \bibinfo {pages} {1679–1696} (\bibinfo {year} {2016})}\BibitemShut {NoStop}%
\bibitem [{\citenamefont {Shoemake}(1985)}]{Shoemake1985}%
  \BibitemOpen
  \bibfield  {author} {\bibinfo {author} {\bibfnamefont {K.}~\bibnamefont {Shoemake}},\ }in\ \href {\doibase 10.1145/325334.325242} {\emph {\bibinfo {booktitle} {Proceedings of the 12th annual conference on Computer graphics and interactive techniques}}},\ \bibinfo {series and number} {SIGGRAPH 85}\ (\bibinfo  {publisher} {ACM Press},\ \bibinfo {year} {1985})\ p.\ \bibinfo {pages} {245–254}\BibitemShut {NoStop}%
\bibitem [{\citenamefont {Sorkine}\ and\ \citenamefont {Alexa}(2007)}]{Sorkine}%
  \BibitemOpen
  \bibfield  {author} {\bibinfo {author} {\bibfnamefont {O.}~\bibnamefont {Sorkine}}\ and\ \bibinfo {author} {\bibfnamefont {M.}~\bibnamefont {Alexa}},\ }\href {10.1145/1281991.1282006} {\bibfield  {journal} {\bibinfo  {journal} {Proceedings of the 5th Eurographics Symposium on Geometry Processing}\ ,\ \bibinfo {pages} {109}} (\bibinfo {year} {2007})}\BibitemShut {NoStop}%
\bibitem [{\citenamefont {Nguyen}, \citenamefont {Jaillet},\ and\ \citenamefont {Redon}(2017)}]{Nguyen2017}%
  \BibitemOpen
  \bibfield  {author} {\bibinfo {author} {\bibfnamefont {M.~K.}\ \bibnamefont {Nguyen}}, \bibinfo {author} {\bibfnamefont {L.}~\bibnamefont {Jaillet}}, \ and\ \bibinfo {author} {\bibfnamefont {S.}~\bibnamefont {Redon}},\ }\href {\doibase 10.1007/s10822-017-0012-y} {\bibfield  {journal} {\bibinfo  {journal} {J. Comput.-Aided Mol. Des.}\ }\textbf {\bibinfo {volume} {31}},\ \bibinfo {pages} {403–417} (\bibinfo {year} {2017})}\BibitemShut {NoStop}%
\bibitem [{\citenamefont {Nguyen}, \citenamefont {Jaillet},\ and\ \citenamefont {Redon}(2018)}]{Nguyen2018}%
  \BibitemOpen
  \bibfield  {author} {\bibinfo {author} {\bibfnamefont {M.~K.}\ \bibnamefont {Nguyen}}, \bibinfo {author} {\bibfnamefont {L.}~\bibnamefont {Jaillet}}, \ and\ \bibinfo {author} {\bibfnamefont {S.}~\bibnamefont {Redon}},\ }\href {\doibase 10.1007/s10822-018-0137-7} {\bibfield  {journal} {\bibinfo  {journal} {J. Comput.-Aided Mol. Des.}\ }\textbf {\bibinfo {volume} {32}},\ \bibinfo {pages} {853–867} (\bibinfo {year} {2018})}\BibitemShut {NoStop}%
\bibitem [{\citenamefont {Hoja}, \citenamefont {List},\ and\ \citenamefont {Boese}(2024)}]{Hoja2023}%
  \BibitemOpen
  \bibfield  {author} {\bibinfo {author} {\bibfnamefont {J.}~\bibnamefont {Hoja}}, \bibinfo {author} {\bibfnamefont {A.}~\bibnamefont {List}}, \ and\ \bibinfo {author} {\bibfnamefont {A.~D.}\ \bibnamefont {Boese}},\ }\href {\doibase 10.1021/acs.jctc.3c01082} {\bibfield  {journal} {\bibinfo  {journal} {J. Chem. Theory Comput.}\ }\textbf {\bibinfo {volume} {20}},\ \bibinfo {pages} {357–367} (\bibinfo {year} {2024})}\BibitemShut {NoStop}%
\bibitem [{\citenamefont {List}, \citenamefont {Boese},\ and\ \citenamefont {Hoja}(2026)}]{List2026}%
  \BibitemOpen
  \bibfield  {author} {\bibinfo {author} {\bibfnamefont {A.}~\bibnamefont {List}}, \bibinfo {author} {\bibfnamefont {A.~D.}\ \bibnamefont {Boese}}, \ and\ \bibinfo {author} {\bibfnamefont {J.}~\bibnamefont {Hoja}},\ }\href {\doibase 10.1063/5.0319348} {\bibfield  {journal} {\bibinfo  {journal} {J. Chem. Phys.}\ }\textbf {\bibinfo {volume} {164}},\ \bibinfo {pages} {194111} (\bibinfo {year} {2026})}\BibitemShut {NoStop}%
\bibitem [{\citenamefont {Mezei}, \citenamefont {Csonka},\ and\ \citenamefont {Kállay}(2015)}]{Mezei2015}%
  \BibitemOpen
  \bibfield  {author} {\bibinfo {author} {\bibfnamefont {P.~D.}\ \bibnamefont {Mezei}}, \bibinfo {author} {\bibfnamefont {G.~I.}\ \bibnamefont {Csonka}}, \ and\ \bibinfo {author} {\bibfnamefont {M.}~\bibnamefont {Kállay}},\ }\href {\doibase 10.1021/acs.jctc.5b00223} {\bibfield  {journal} {\bibinfo  {journal} {J. Chem. Theory Comput.}\ }\textbf {\bibinfo {volume} {11}},\ \bibinfo {pages} {2879–2888} (\bibinfo {year} {2015})}\BibitemShut {NoStop}%
\end{thebibliography}
\end{document}